\documentclass[arxiv,article,moreauthors,pdftex]{mdpimod}

\firstpage{1}
\issuenum{1}
\articlenumber{5}
\pubyear{2020}
\copyrightyear{2020}
\history{Received: date; Accepted: date; Published: date}

\newcommand{\avg}[1]{{\langle #1 \rangle}}
\newcommand{\ds}{\displaystyle}

\Title{Evolution of Cooperation in the Presence of Higher-Order Interactions: from Networks to Hypergraphs
}

\Author{\orcidA{}Giulio Burgio $^{1}$, \orcidB{}Joan T. Matamalas $^{2}$, \orcidC{}Sergio G\'omez $^{1}$ and \orcidD{}Alex Arenas $^{1}$*}

\AuthorNames{Giulio Burgio, Joan T.\ Matamalas, Sergio G\'omez and Alex Arenas}

\address{%
$^{1}$ \quad Departament d'Enginyeria Inform\`atica i Matem\`atiques, Universitat Rovira i Virgili, Av. Pa\"isos Catalans 26, Tarragona 43007, Spain; E-Mails: giulioburgio@gmail.com (G.B.), sergio.gomez@urv.cat (S.G.)\\
$^{2}$ \quad Harvard Medical School \& Brigham and Women's Hospital, 75 Francis St, Boston MA 02115, USA; E-mail: jmatamalas@bwh.harvard.edu}

\corres{Author to whom correspondence should be addressed; E-Mail: alexandre.arenas@urv.cat}

\abstract{Many real systems are strongly characterized by collective cooperative phenomena whose existence and properties still need a satisfactory explanation. Coherently with their collective nature, they call for new and more accurate descriptions going beyond pairwise models, such as graphs, in which all the interactions are considered as involving only two individuals at a time. Hypergraphs respond to this need, providing a mathematical representation of a system allowing from pairs to larger groups. In this work, through the use of different hypergraphs, we study how group interactions influence the evolution of cooperation in a structured population, by analyzing the evolutionary dynamics of the public goods game. Here we show that, likewise network reciprocity, group interactions also promote cooperation. More importantly, by means of an invasion analysis in which the conditions for a strategy to survive are studied, we show how, in heterogeneously-structured populations, reciprocity among players is expected to grow with the increasing of the order of the interactions. This is due to the heterogeneity of connections and, particularly, to the presence of individuals standing out as hubs in the population. Our analysis represents a first step towards the study of evolutionary dynamics through higher-order interactions, and gives insights into why cooperation in heterogeneous higher-order structures is enhanced. Lastly, it also gives clues about the co-existence of cooperative and non-cooperative behaviors related to the structural properties of the interaction patterns.
}


\begin{document}

\section{Introduction}
\label{sec:intro}

In behavioral terms, cooperation is the providing of a benefit to another individual at some cost for the provider. In well-mixed populations, where all individuals interact with each other, evolutionary game theory predicts that cooperation cannot survive due to the existence of selfish behaviors \cite{roca1,szabo}. Nevertheless, we do observe cooperation in many different real systems, ranging from genomes to human societies \cite{dawkins,pennisi,dugatkin,nowak}, and not as a marginal phenomenon: rather, it often strongly shapes and makes even possible the existence of those systems. For example, without cooperative behaviors like the work division and the establishment and respect of social norms, the humankind could not have developed even the simplest forms of society. Among microbes, some viruses and cells have been observed to be involved in (involuntary) cooperative behaviors that are essential for their reproduction and diffusion \cite{pennisi,nowak}. The existence of cooperation in biological and social systems has led to hypothesize several mechanisms to explain its ubiquity: inclusive fitness, direct and indirect reciprocity, punishment, cultural group selection, etc. (clearly, those requiring substantial cognitive demands, apply only to human's social systems) \cite{hamilton,schneider,trivers,nowak2,bshary,gintis,coyne}. Among those mechanisms, we focus on the so-called \textit{network reciprocity} \cite{nowak3,ohtsuki,killingback,santos,gomez-gardenes,santos2,gomez-gardenes2,tanimoto,traulsen}.

Generally speaking, an act of reciprocity takes place when some individual returns a beneficial act received by someone else, establishing a fruitful mutual exchange. Therefore, it is a cooperative act. In particular, one refers to network reciprocity when the mutual cooperative exchange is sustained by the existence of a suitable structure of interactions, represented through a network, among the individuals of a population. Two conditions lie at the base of network reciprocity: a limited, recurrent set of interacting game opponents (thus avoiding the anonymity between them); and a local adaptation mechanism through which an individual can change (update) its behavior (strategy) depending on how its neighbors in the network behave and perform. Apart from its fascinating theoretical character, network reciprocity is regarded as one of the most important and interesting of the mechanisms from an application point of view. Indeed, the easiness of its requirements to be fulfilled, makes it a transverse mechanism applicable not only to those system made of rational units (e.g., social, economical), but also to those whose units are capable of only very simple behaviors (e.g., many biological systems) \cite{tanimoto}.

The use of networks to formalize the topology of interactions within a system, however, tacitly assumes that all the interactions that occur in the system are essentially pairwise. This may not be a satisfactory description for those real systems exhibiting higher-order interactions to which more than two entities participate together and which are not reducible to combinations of lower-order ones. Examples of such interactions include triadic \cite{granovetter} and higher-order closures in social networks \cite{matamalas}, co-authorship networks in science \cite{patania}, spatial coexistence relations among species in an ecosystem \cite{levine}, and trigenic interactions in genetic networks \cite{kuzmin}. In all such cases, representing the interaction patterns through a network (that is, approximating those higher-order interactions with pairwise ones) implies the loss of essential information to understand their dynamics \cite{lambiotte}.

This work tries to bring network reciprocity closer to reality. We investigate the effect of considering higher-order interactions, using hypergraphs, on the evolution of cooperation when the population of individuals possesses structural correlations. Hypergraphs allow to distinguish the dynamics on a clique of $m$ individuals (an $m$-clique), in which each of the individuals (vertices) interacts separately, in pair, with each of the other $m$$-$$1$ individuals, from the dynamics in which the $m$ individuals interact all together, as a group, in an intrinsically different way (represented as a hyperedge of cardinality $m$). Moreover, hypergraphs, and also other higher-order representations as well \cite{matamalas2}, solve the lack of uniqueness in defining group interactions from pairwise information only.

We show here that the mechanism of network reciprocity can be successfully extended to higher-order structures and that, in fact, it even becomes reinforced.

\section{Dynamics}
\label{sec:dynamics}
\unskip
\subsection{Public Goods Dilemma}
\label{sec:pgg}

To study cooperation in presence of higher-order interactions we need a game definable for any number of players (the size of the group) and belonging to the class of social dilemmas. In such games, the equilibrium solution under unilateral changes of strategy (i.e., considering changes of strategy of only one player at a time), called Nash equilibrium, is not the most efficient solution for all the players (i.e., it is not Pareto-efficient). In particular, in the type of game we are interested in, the equilibrium solution is the one in which all the players choose to defect, although the Pareto-efficient solution is the one in which they all cooperate, hence the social dilemma. The prototypical $m$-players dilemma of this kind is the \textit{public goods game} (PGG). Each player, independently, chooses whether to put (cooperating) or not (defecting) a certain amount $b$ (fixed for all players, in our setup) into a public pot. The collected sum is re-scaled by a \textit{synergy factor} $\alpha\in\mathbb{R}$, $\alpha>1$, and equally redistributed among the $m$ players disregarding of their strategy. When $n_c\leq m-1$ neighbors cooperate, a player with strategy $\sigma$ (set to 1 if cooperates, 0 if defects) gets a payoff $f(\sigma)$ given by
\begin{equation}
  f\left(\sigma\right) = b\left[\left(\frac{\alpha}{m}-1\right)\sigma+\frac{\alpha\left(m-1\right)}{m}q\right]
  \label{payoff}
\end{equation}
where $q\equiv n_c/(m-1)$ is the fraction of neighbors that cooperate. The first term accounts for the cost and proportional gain from the contribution to the pot of the player, while the second corresponds to the gain received from the cooperating neighbors. Since the second term is always non-negative, the best response for a player depends on whether $\alpha$ is smaller or greater than $m$. The dilemma exists for $\alpha<m$ because, in this case, according to Equation~(\ref{payoff}), the best strategy for any player is to defect. However, each of them gets a null payoff, whereas it would be positive if all of them chose to cooperate.

\unskip
\subsection{Structural Connectivity}
\label{sec:structure}

We give here a brief introduction to the connectivity structures we use in this work: simple hypergraphs. Given a finite set $V$ of vertices and a family $E$ of subsets $\{e_I\}_{I\subseteq V}$ of $V$, the couple $H=(V,E)$ defines a \textit{hypergraph}, and each $e_I$ is called \textit{hyperedge}, representing a relation among all the vertices in it. The maximum and minimum cardinality (degree) of the hyperedges in $H$ are called the \textit{rank} and \textit{co-rank} of the hypergraph, respectively: $\mbox{rank}(H)=m_{\max}$ and $\mbox{co-rank}(H)=m_{\min}$.
If they both are equal, with value $s\in\mathbb{N}$, the hypergraph is said to be \textit{$s$-uniform} \cite{bretto}.

Another important concept is the \textit{$2$-section} of $H$. We can define it as the graph whose vertices are the vertices of $H$, and where two distinct vertices form an edge if and only if they belong to, at least, one common hyperedge in $H$ \cite{bretto}.

Additionally, an hypergraph $H$ is called \textit{simple} if $e_I\subseteq e_J\Rightarrow I = J$. This means that it has no repeated hyperedges, and can be obtained from a generic hypergraph $(V,E)$ removing from $E$ all the hyperedges which are subsets of at least another hyperedge.

Finally, given the family $S(i)$ of hyperedges $\{e_J\}_{J\subseteq V}$ containing vertex $i$, and assuming no repeated hyperedge, we define the \textit{$m$-degree} $k_{m}^{(i)}$ of $i$ as the number of hyperedges of degree $m$ in $S(i)$. The generalized degree of node $i$ is the number of hyperedges incident on it, given by
\begin{equation}
  k^{(i)} = \sum_{m=m_{\min}}^{m_{\max}} k_{m}^{(i)}
\end{equation}

We construct a hypergraph starting from a certain base network and converting in hyperedges a randomly selected fraction~$p$ of its 3-cliques, see Section~\ref{sec:algorithm} for further details. From now on, we will call these selected hyperedges as \textit{triangles}.

\unskip
\subsection{Public Goods Dynamics in Hypergraphs}
\label{sec:model}

At the base of our model lie the following microscopic rules, imposed at each time step (round): (i) each player plays with its strategy, either cooperate or defect, within each of the hyperedges incident on it; (ii) each strategist chooses one of its neighbors uniformly at random in the 2-section of the hypergraph, and performs the strategy update. These steps are performed synchronously by all the players.

The payoff $f^{(i)}(t)$ that player $i$ with $m$-degree $k_{m}^{(i)}$ and playing with strategy $\sigma_i(t)$ got at the end of a time step $t$, obeys the following dynamic equation:
\begin{equation}
  f^{(i)}(t) = b\sum_{m=m_{\min}}^{m_{\max}} k_{m}^{(i)} \left[\left(\frac\alpha m-1\right)\sigma_i(t) + \frac{\alpha(m-1)}{m} q_{m}^{(i)}(t) \right]
\label{payoff_dynamics}
\end{equation}
where $q_{m}^{(i)}(t)$ is the probability that a neighbor of node~$i$, in a hyperedge of degree $m$, cooperates at time $t$.

To explore different selection pressure regimes, we consider \textit{smoothed imitation} by taking, as probability function for the strategy update, the Fermi distribution function:
\begin{equation}
  {\cal F}(\Delta f;\beta) = \frac{1}{1+e^{-\beta \Delta f}}
\end{equation}
where $\Delta f$ is the difference between the payoff of the randomly chosen neighbor and the payoff of the focal player, and parameter $\beta > 0$ allows to regulate the strength of the selection process: strong and sharp for high values of $\beta$, weak and noisy for low values).

The global state of a population of size $N$ is defined by the density of cooperators:
\begin{equation}
  c(t) = \frac{1}{N}\sum_{i=1}^N \sigma_i(t)
\end{equation}
We initialize the system with a fraction $c_0\equiv c(t=0)$ of randomly selected cooperators.

From Equation~(\ref{payoff_dynamics}) it immediately follows that $c=0$ and $c=1$ are absorbing states of the dynamics, for no change of state is possible from them.

\section{Results}
\label{sec:results}

For every set of hypergraphs we report the effective asymptotic cooperators density $c_\infty$ and the convergence time $t$ as a function of $\alpha$ and for different values of the conversion fraction $p$, as found in the performed Monte Carlo (MC) simulations; see Section~\ref{sec:mc_simu} for further details. To systematically reveal the effect on the dynamics of converting 3-cliques in triangles, we report the results found for hypergraphs generated through the Holme-Kim model (HK) \cite{kim} and the Dorogotsev-Mendes model (DM) \cite{dorogotsev}. They both grow heterogeneous networks having topological properties typical of many real complex systems, and in particular social networks, like power-law degree distributions, the small-world property, and high transitivity (i.e., high clustering coefficient).

Let us call $\alpha_{\textup{surv}}(p)$ the value of $\alpha$ at which, for a given $p$, the asymptotic cooperator density $c_\infty$ starts to grow above zero, to be compared with the corresponding critical value $\alpha_{\textup{cr}}(p)$ for a well-mixed population. To this purpose, we define
\begin{equation}
  r(p) \equiv \frac{\alpha_{\textup{surv}}(p)}{\alpha_{\textup{cr}}(p)}
\end{equation}
where
\begin{equation}
  \alpha_{\textup{cr}}(p) = \sum_{m=m_{\min}}^{m_{\max}} m\,p_m
\end{equation}
being $p_m$ the fraction of hyperedges of cardinality~$m$ in the hypergraph. Values of $r(p)$ smaller than 1 mean some level of structural reciprocity favoring the survival of cooperation. As a particular case, $r(0)$ measures the network reciprocity.

\begin{figure}[tb!]
  \centering
  \includegraphics[width=.46\linewidth]{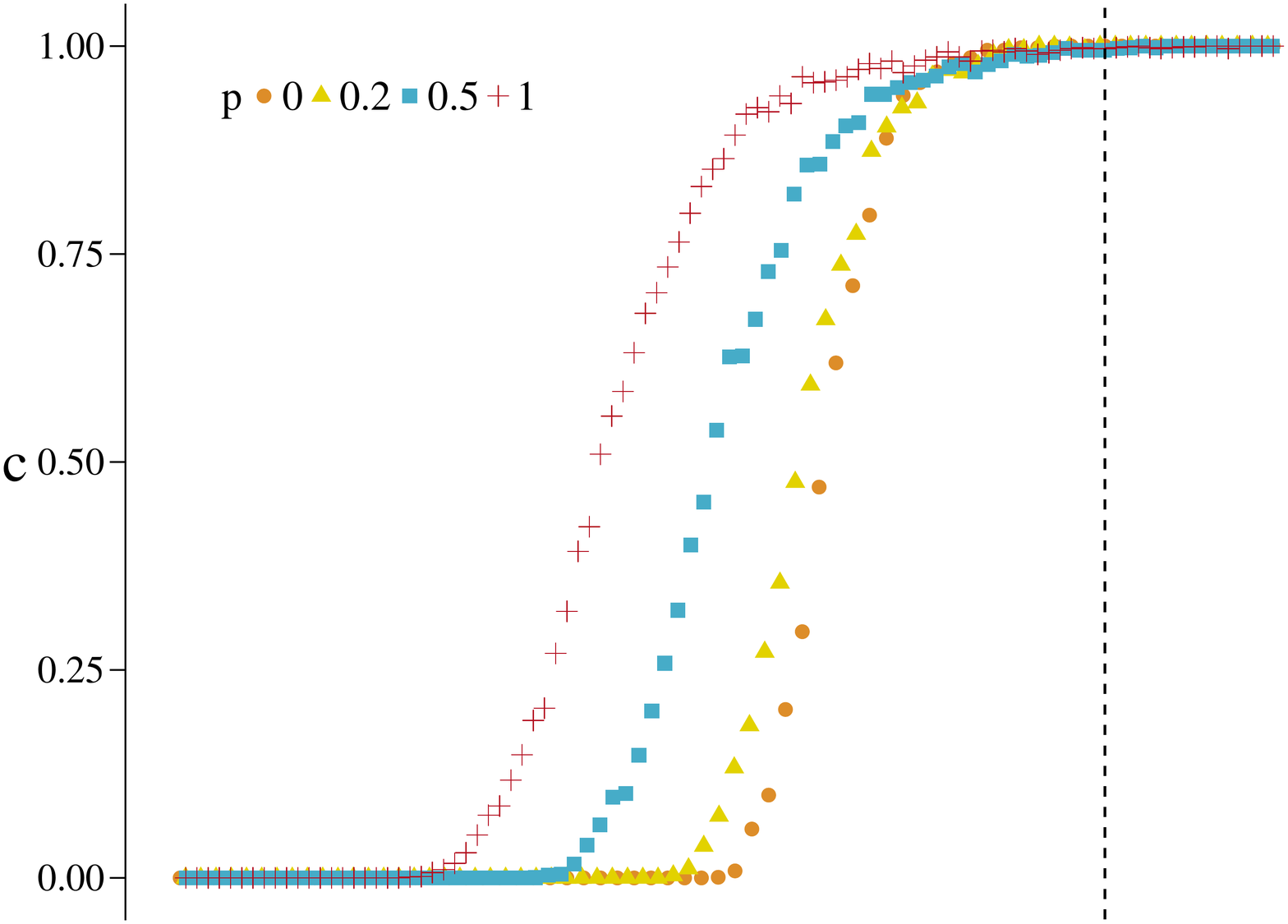}
  \hspace{0.5cm}
  \includegraphics[width=.46\linewidth]{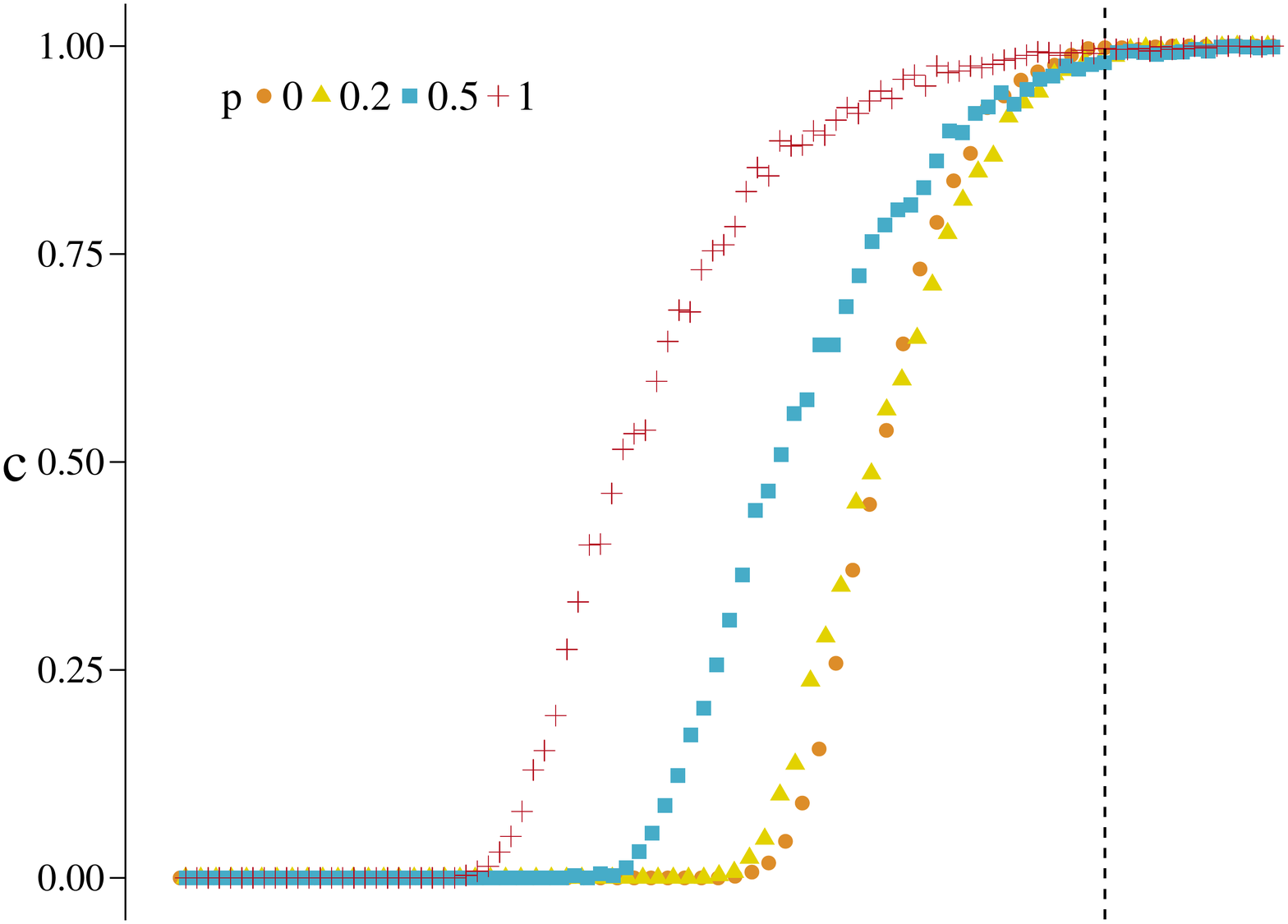}
  \\
  \includegraphics[width=.46\linewidth]{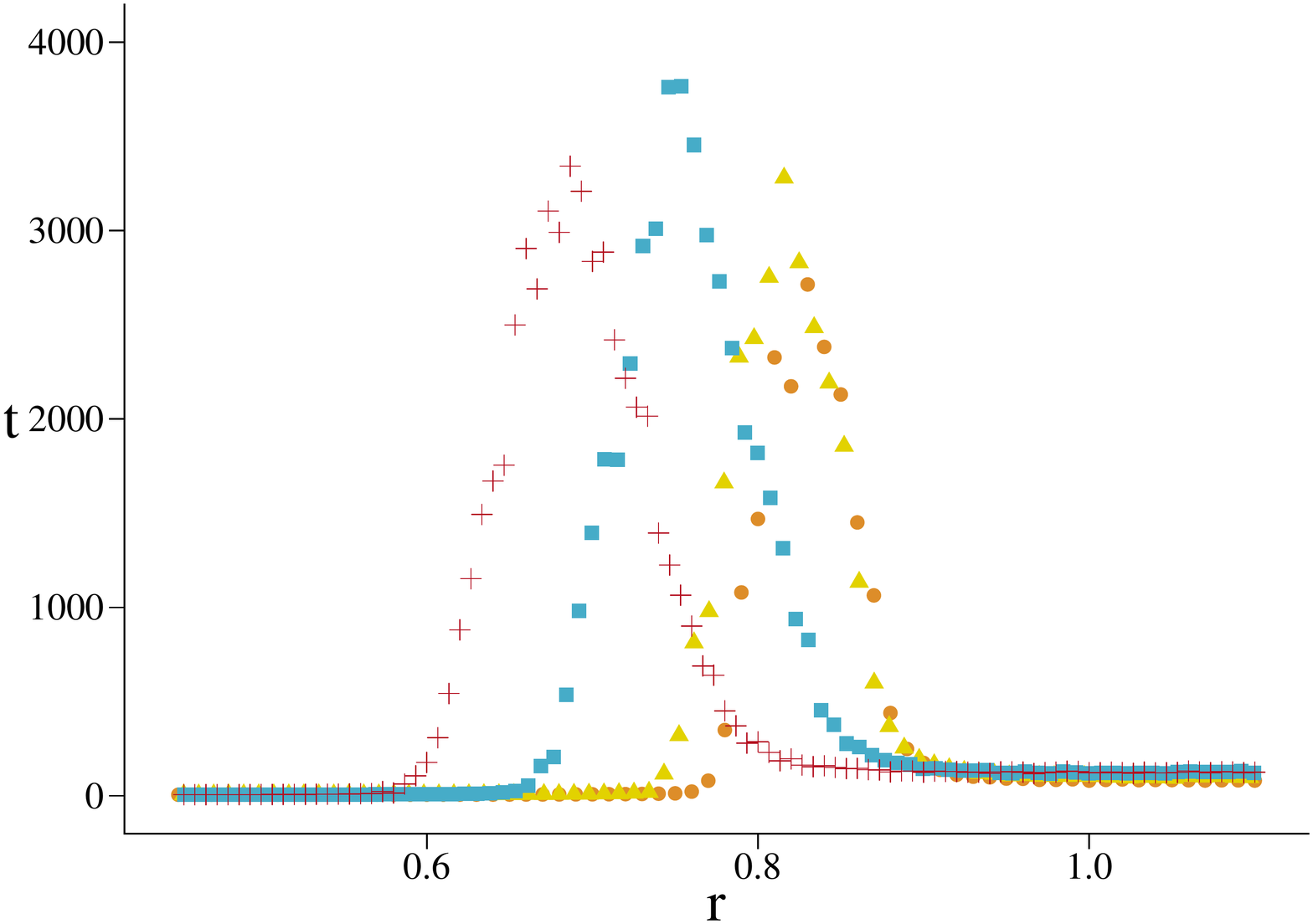}
  \hspace{0.5cm}
  \includegraphics[width=.46\linewidth]{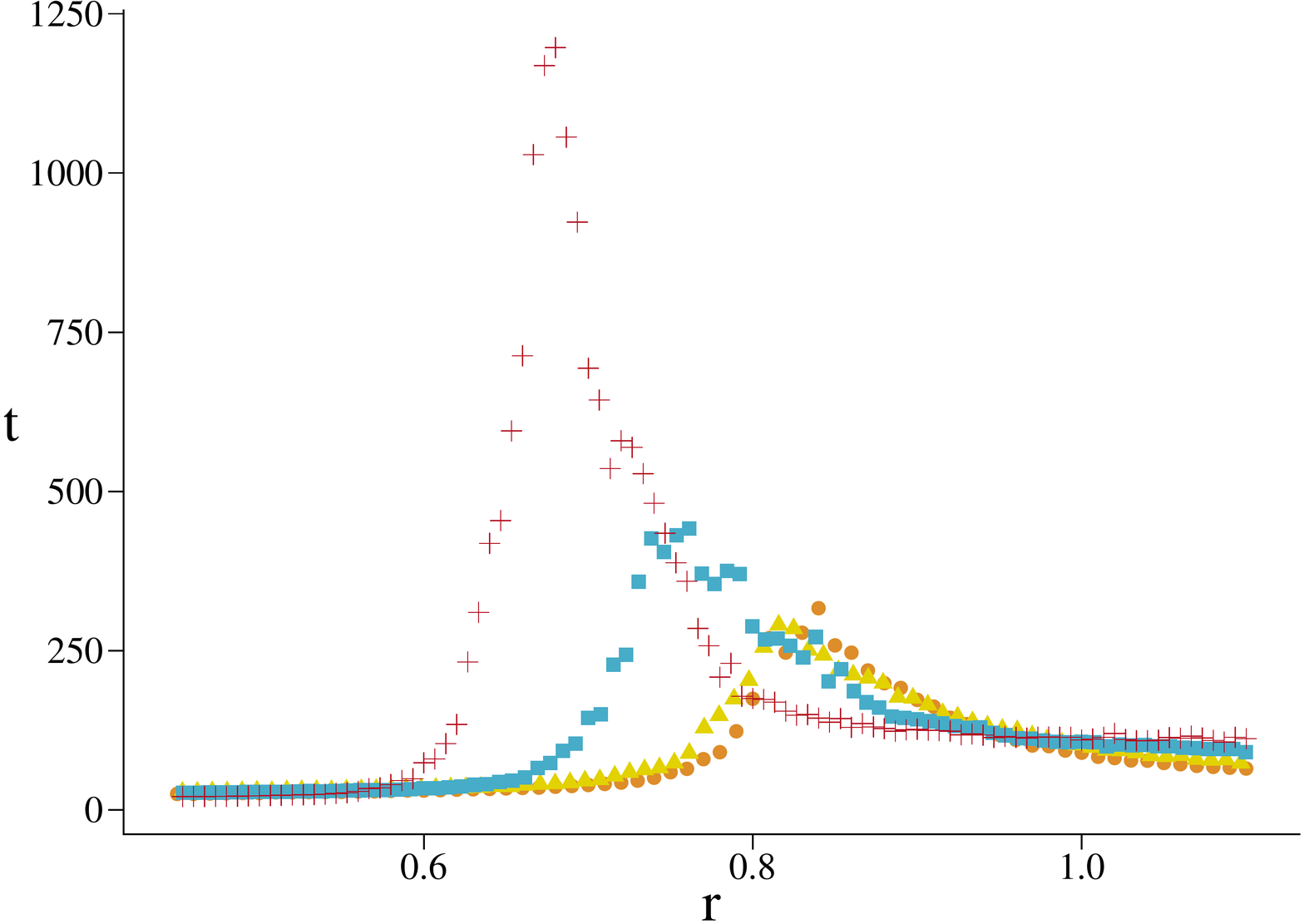}
  \caption{Asymptotic density of cooperators $c$ (top) and convergence time $t$ (bottom) versus $r$, for hypergraphs generated from HK networks with $N=500$ vertices, $m=3$ (average degree $\langle k\rangle\approx 6$), $P_t=1$, and for $\beta=1$ (left) and $\beta=0.1$ (right). The values of $p$ are specified in the legend, and $c_0=0.5$. The dashed line at $r=1$ indicates the abrupt transition for a well-mixed population, as predicted by the MF theory.}
\label{HK}
\end{figure}

In Figure~\ref{HK} we show the results for MC simulations with $c_0=0.5$ and two values of~$\beta$, on hypergraphs stemming from HK networks with $N=500$ vertices, $m=3$ (average degree $\langle k\rangle \approx 6$ on the 2-section) and $P_t=1$. Parameter $P_t$, the probability of making a triad-formation step after a preferential-attachment step in the HK model, has been set to~1 to get the maximum possible transitivity, thus making more evident the effects of increasing the fraction~$p$ of converted 3-cliques. We find that, as we increase~$p$ from~0 to~1, $r(p)$ decreases monotonically from~0.78 to~0.60 when $\beta=1$, and from~0.79 to~0.63 when $\beta=0.1$. Thus, reciprocity is enhanced when we increase~$p$, the fraction of 3-cliques promoted to triangular hyperedges. We also note how, by decreasing $\beta$ (weaker selection), the structural reciprocity decreases as well.

Looking at the convergence times in Figure~\ref{HK}, we see the typical peaked shape characterizing the critical region. Dealing with small finite systems, we expect the convergence time to show a peak near the transition, but still to remain finite. Since the maximum number of allowed time steps $t_{\max}$ is fixed to $2\times10^4$, and we average over 100~realizations, we can state that, if the average convergence time is smaller than $200$, then all the trials converged and the final populations are perfectly homogeneous; note that this is only a sufficient condition. In this case, an effective cooperation density $c_\infty$ other than~0 or~1 represents an average of homogeneous final states. This is the case for $r$~values out of the transition region.

Counting how many times the system has not converged $n_{NC}$ in 100~realizations, and knowing the value of the average convergence time~$t$, the average time $t_C$ the system took to converge in the remaining $100-n_{NC}$ trials is $t_C=100[t-n_{NC}(t_{\max}/100)]/(100-n_{NC})$. For the previous hypergraphs we find $n_{NC}\leq12.3$ for $\beta=1.0$ and $n_{NC}\leq2.4$ for $\beta=0.1$, in the critical regions. The maximum values of $t_C$ are found to be around 1490 and 710, respectively. Being $t_C$ well below $t_{\max}$, we expect that the non-convergence of some realizations does not depend on the value chosen for the latter. This suggests that the lack of convergence is not a purely dynamical effect ---like at the critical point for an infinite well-mixed population---, but it is due to the topological constraints of the structure. In fact, we find the presence of configurations of local states preventing the system to escape from them, acting as \textit{topological traps}, see Section~\ref{sec:traps}.

We also note that with the increasing of $p$, the critical region gets wider, whatever is the value of $\beta$. In Section~\ref{sec:isolated_defector}, we give an explanation for this based on the degree heterogeneity and the local clustering of the structure.

\begin{figure}[tb!]
  \centering
  \includegraphics[width=.46\linewidth]{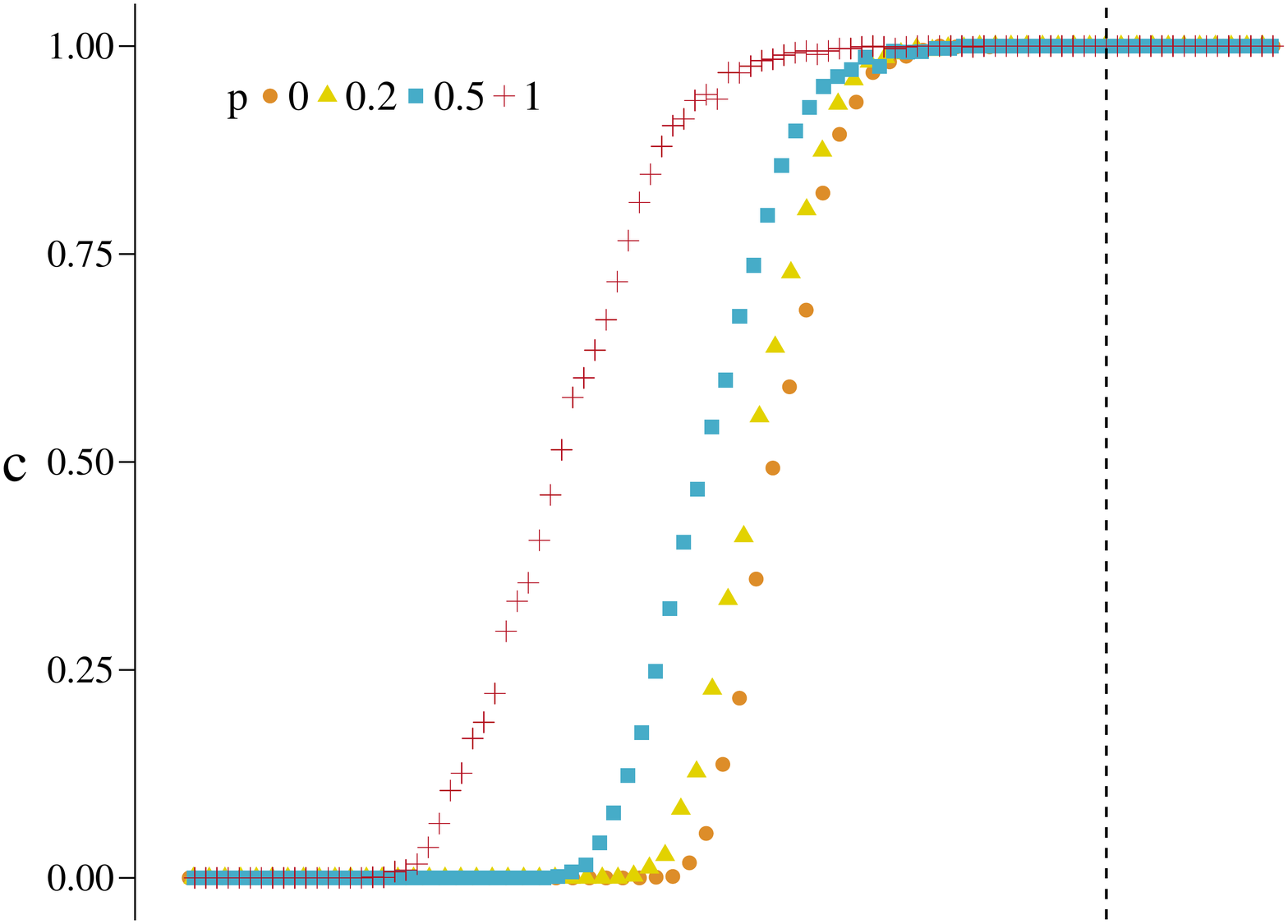}
  \hspace{0.5cm}
  \includegraphics[width=.46\linewidth]{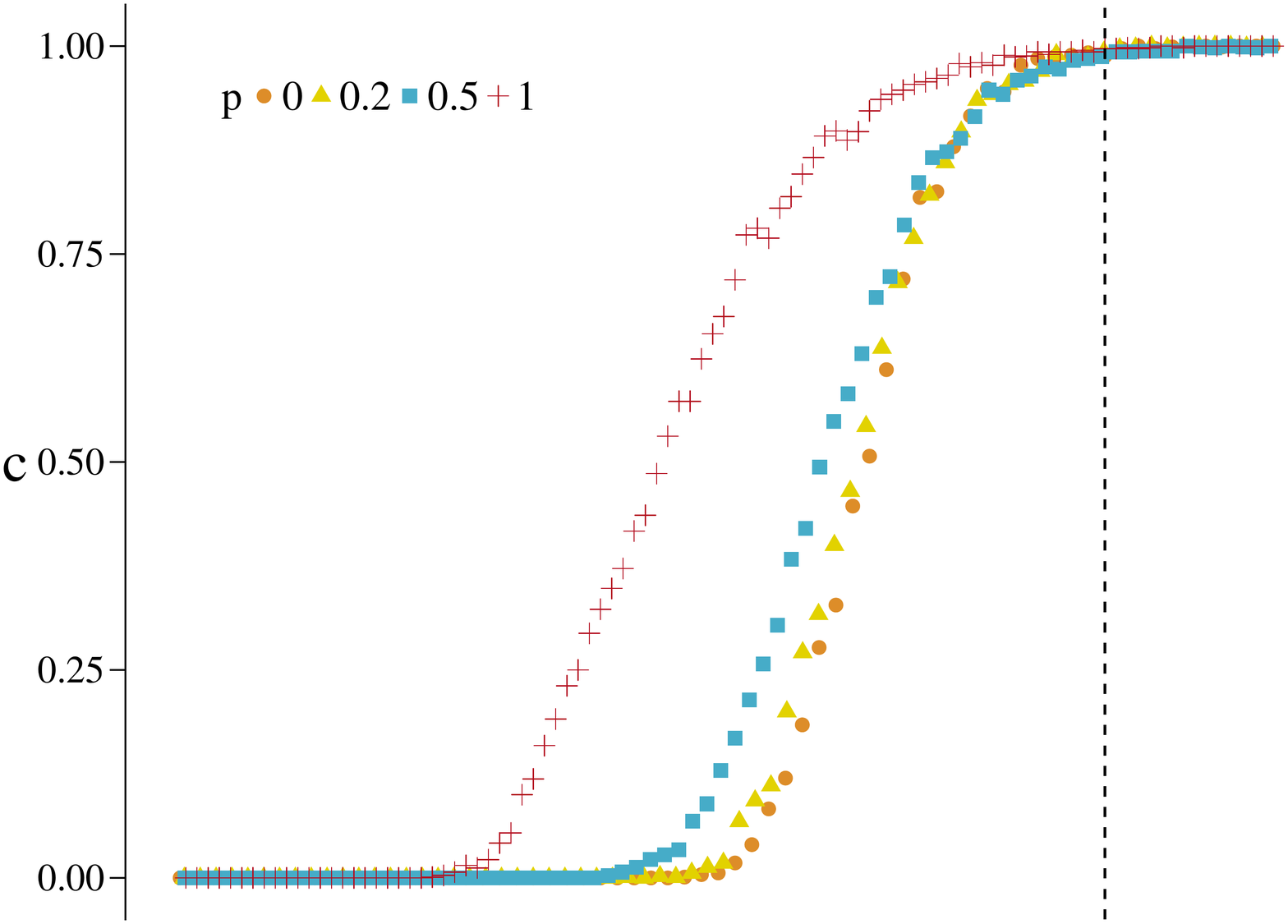}
  \\
  \includegraphics[width=.46\linewidth]{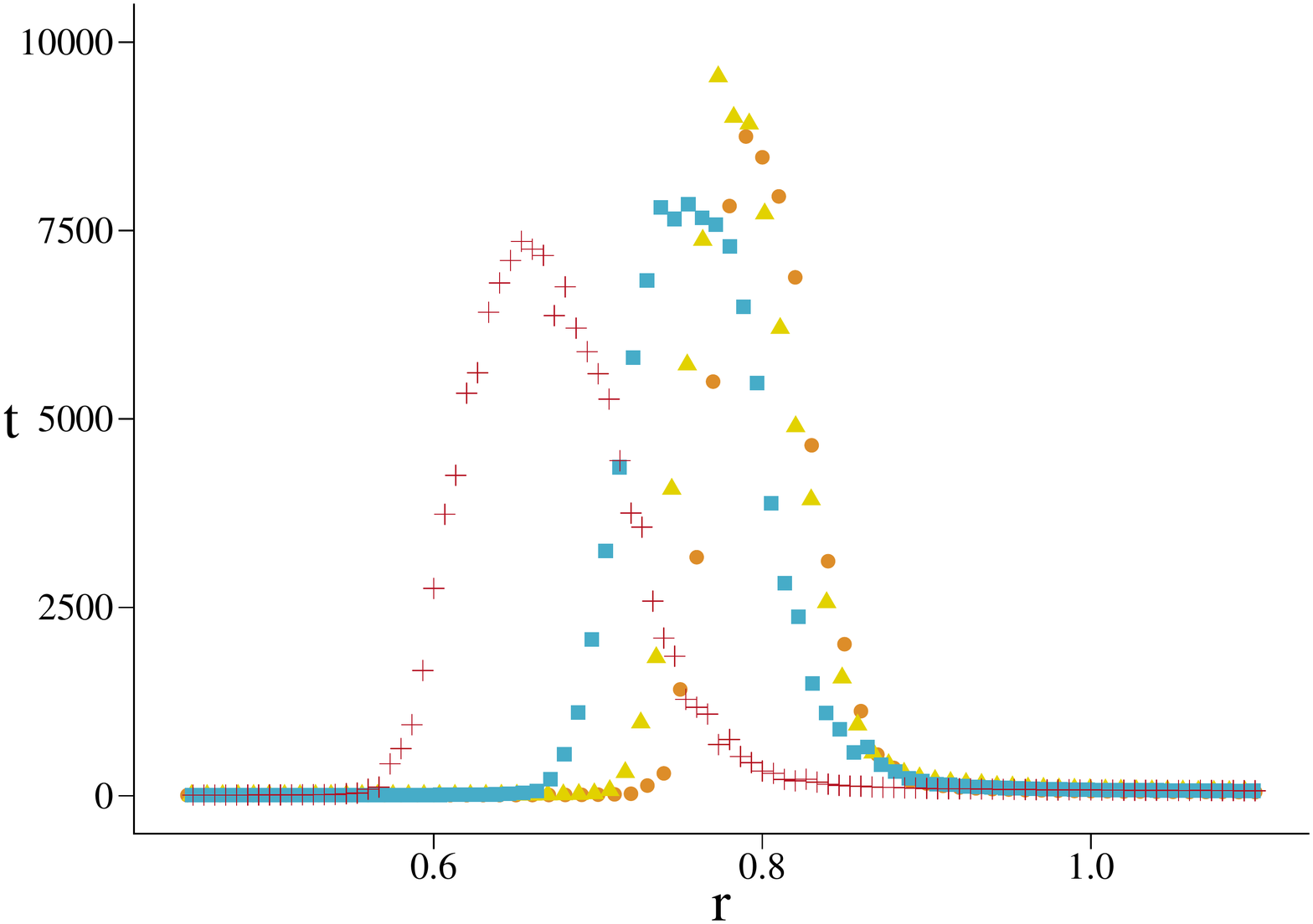}
  \hspace{0.5cm}
  \includegraphics[width=.46\linewidth]{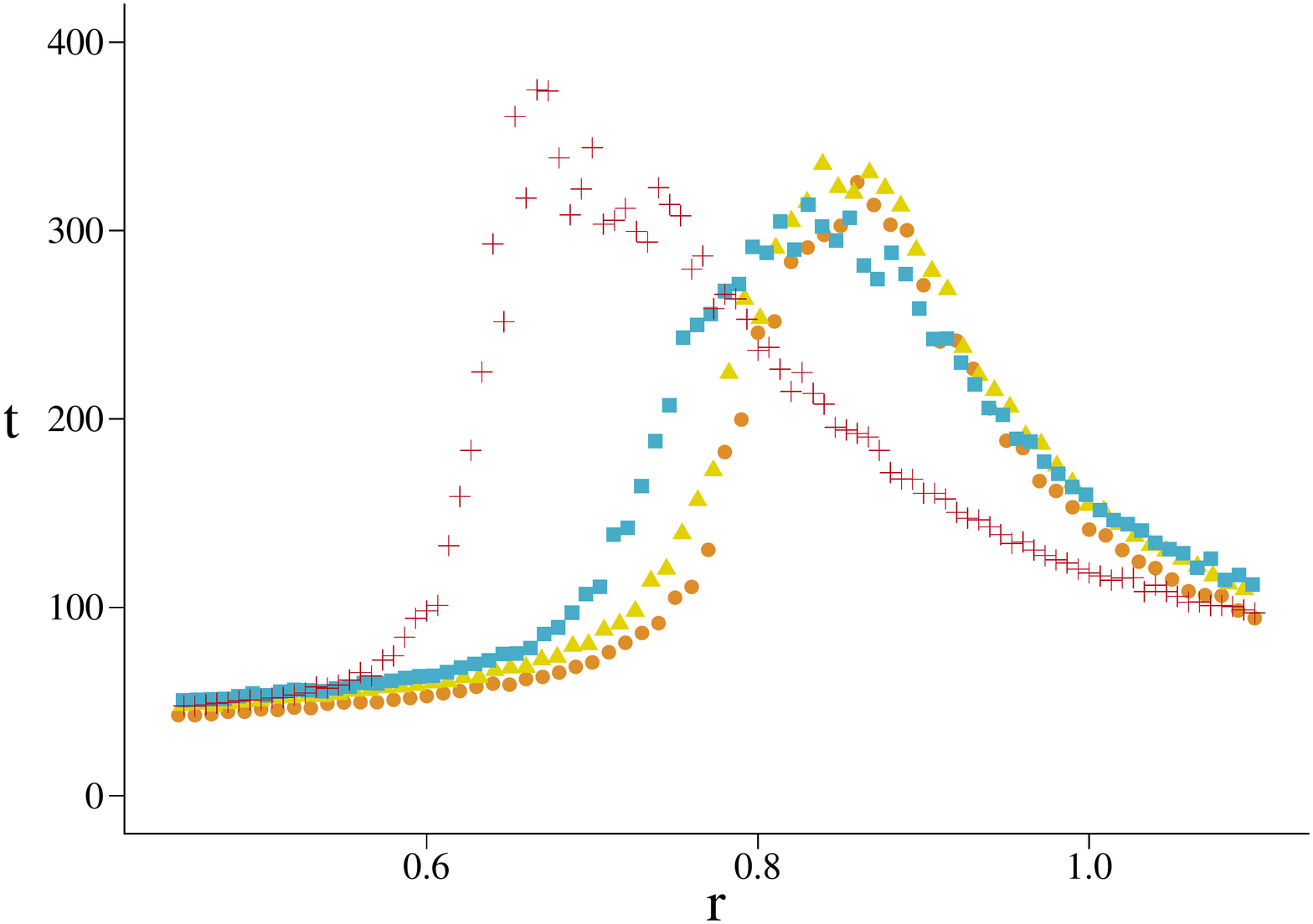}
  \caption{Asymptotic density of cooperators $c$ (top) and convergence time $t$ (bottom) versus~$r$, for hypergraphs generated from DM networks with $N=500$ vertices (average degree $\langle k\rangle=4$) and for $\beta=1$ (left) and $\beta=0.1$ (right). The values of~$p$ are specified in the legends, and $c_0=0.5$. The dashed line at $r=1$ indicates the first-order transition for a well-mixed population, as predicted by the MF theory.}
\label{DM}
\end{figure}

Quite similar results are found for hypergraphs stemming from DM networks. In Figure~\ref{DM} we show the results obtained for base networks with $N=500$ vertices ($\langle k\rangle=4$). Going from $p=0$ to $p=1$, we get the following monotonically decreasing values of $r(p)$: from 0.75 to 0.58 for $\beta=1$, from 0.78 to 0.62 for $\beta=0.1$. Furthermore, we find, respectively, $n_{NC}\leq39.7$ and $n_{NC}=0$ in the critical regions. The maximum values of $t_C$ are found to be around 2660 and 380. As for HK hypergraphs, setting $t_{\max}=5\times10^4$, we find no significant differences. In Appendix~\ref{sec:appendix_b} we report the results obtained also for $c_0=0.3$ and $c_0=0.7$.

In the case of hypergraphs derived from Erd\H os-R\'enyi networks (ER) \cite{erdos}, the transitions are always sharp and just below $r=1$ (see Appendix~\ref{sec:appendix_c}), given the high structural homogeneity and low clustering of ER networks. The average convergence time is found not over 300~steps, along with the absence of traps.

\section{Mathematical Analysis}
\label{sec:mathematical_analysis}

\unskip
\subsection{Mean-Field Approximation}
\label{sec:MF}

The mean-field (MF) approach relies on two approximations: (i) all the vertices have the same $m$-degree sequence $(\langle k_{m_{\min}}\rangle,\ldots,\langle k_{m_{\max}}\rangle)$, averaged over the original hypergraph with $m$-degree sequence distribution $P(k_{m_{\min}},\ldots,k_{m_{\max}})$; (ii) the local states of different vertices are not correlated. Then, a simple master equation for the density of cooperators $c(t)$ follows:
\begin{equation}
  c(t+1)=c(t)+c(t)\left(1-c(t)\right)\left({\cal F}(\Delta f)-{\cal F}(-\Delta f)\right)
\label{MF_equation}
\end{equation}
where
\begin{equation}
  \Delta f = \langle f^{(c)}\rangle-\avg{f^{(d)}} = b \sum_{m=m_{\min}}^{m_{\max}} \left(\frac{\alpha}{m}-1\right) \avg{k_m}
\end{equation}
Imposing $c(t+1)=c(t)$ we get, besides the absorbing state solutions $c(t)=0$ and $c(t)=1$, $\forall \alpha$, the stationary solution corresponding to $\mathcal{F}(\Delta f)=\mathcal{F}(-\Delta f)$, that is, $\Delta f=0$, given by
\begin{equation}
  \alpha = \alpha_{cr} \equiv \frac{{\ds\sum_{m=m_{\min}}^{m_{\max}}}\avg{k_m}}{\ds\sum_{m=m_{\min}}^{m_{\max}}m^{-1}\avg{k_m}}=\sum_{m=m_{\min}}^{m_{\max}}mp_m=\avg{m}
  \label{MF_critical_point}
\end{equation}
for all $c(0)\in(0,1)$, being
\begin{equation}
  p_m = \frac{m^{-1}\avg{k_m}}{\ds\sum_{m=m_{\min}}^{m_{\max}}m^{-1}\avg{k_m}}
\end{equation}
the fraction of hyperedges of cardinality~$m$ or, for a well-mixed population, the probability of playing in a group of size $m$, and $\avg{m}$ their average cardinality.  Equation~(\ref{MF_critical_point}) gives the expected abrupt phase transition between full defection and full cooperation, for an infinite well-mixed population, in the presence of either one or more allowed group sizes (corresponding to uniform and non-uniform hypergraphs, respectively).

To be precise, the shown abrupt transition holds in the limit of an infinite population. The finite-size, trivially-structured version of a well-mixed population is an all-connected-to-all structure. Then, considering a $m$-uniform complete simple hypergraph of $N$~vertices, it can be shown (see Appendix~\ref{sec:appendix_a}) that, for $\beta b\gg 1$, the transition is at $\alpha\equiv\alpha_{\textup{cr}}^m=m(N-1)/(N-m)$, greater than $\alpha_{\textup{cr}}=m$ for any finite value of $N$. Additionally, $\alpha_{\textup{cr}}^m/\alpha_{\textup{cr}}$ grows with~$m$. This finite-size effect slightly raises the structural reciprocity herein reported.

\unskip
\subsection{Invasion Analysis}
\label{sec:invasion}

The MF prediction is clearly reliable only for highly homogeneous structures. In all the other cases, instead, it serves as a reference point for measuring the structural reciprocity due to the complex heterogeneity of connections. In this section, we study the conditions for the invasion of the structured population, or for the resistance to be invaded for a player, using a certain strategy in specific configurations. We give a justification to the higher reciprocity level and the enlargement of the transition region, both found when increasing $p$ in the used heterogeneous hypergraphs.

\subsubsection{The Role of Cooperator Hubs}
\label{sec:cooperator_hubs}

For all the analyzed scale-free networks in Figures~\ref{HK} and~\ref{DM}), we observe the transition moving towards notable smaller values of $r$, whatever is the value of $p$. This is not the case for homogeneous hypergraphs, like the ones derived from ER networks. We show here that the key difference is the presence of hubs in the former networks.

\begin{figure}[tb!]
  \centering
  \includegraphics[width=.48\linewidth]{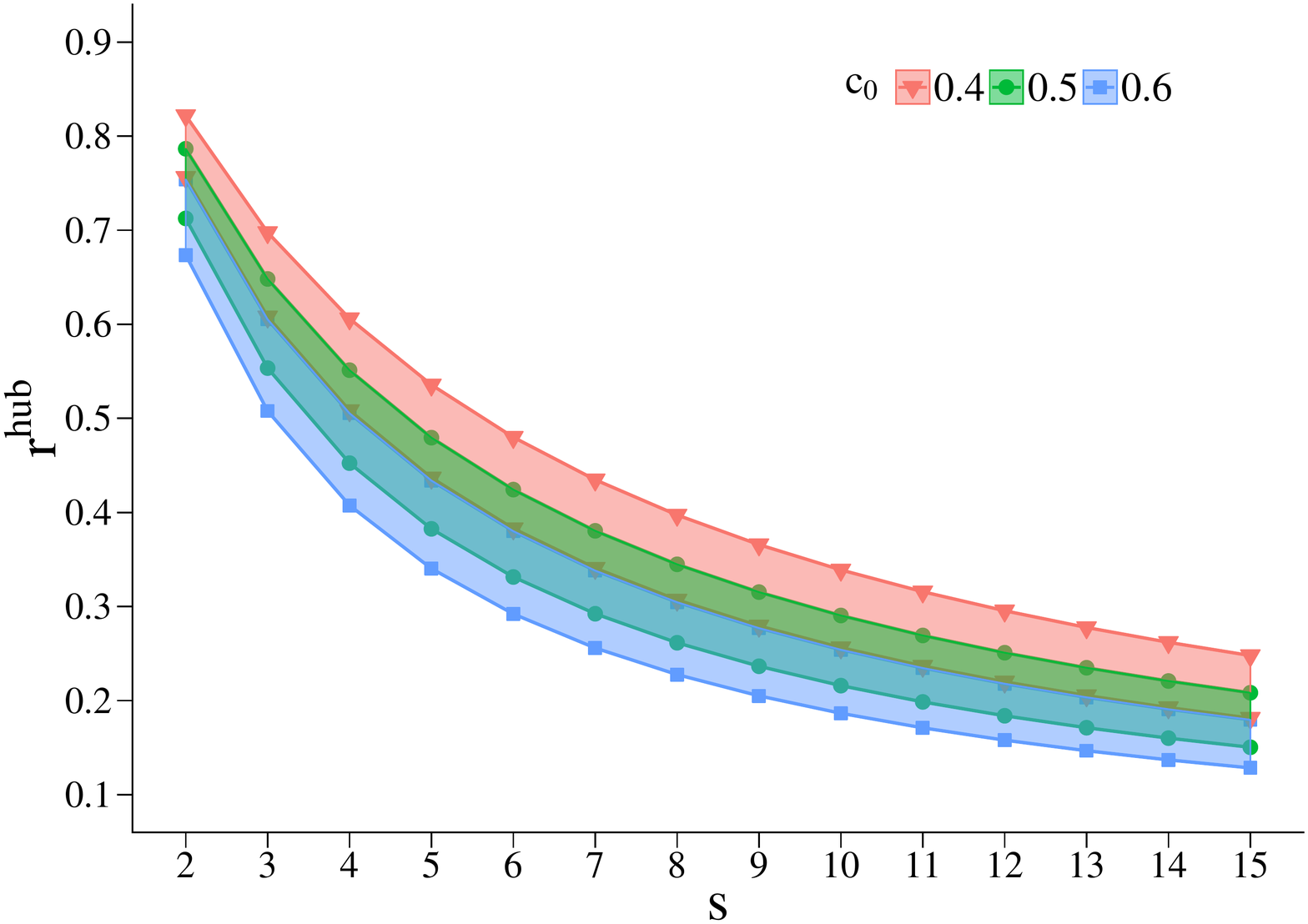}
  \hfill
  \includegraphics[width=.48\linewidth]{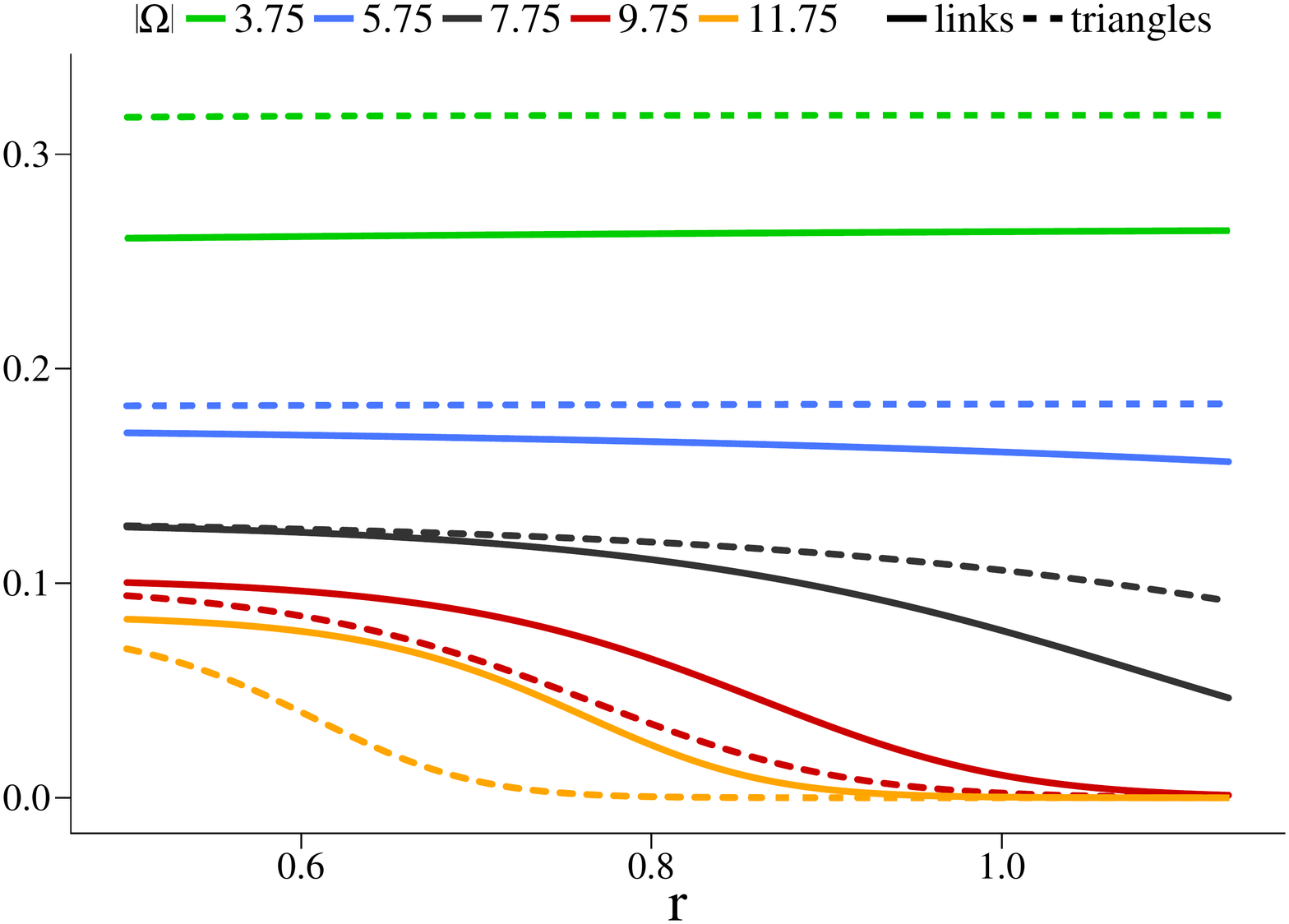}
  \caption{{\em Left}. The ratio $r^{\textup{hub}}\equiv\alpha_{\textup{th}}^{\textup{hub}}/\alpha_{\textup{cr}}$ versus the rank $s$ of a $s$-uniform hypergraph, for the survival of a cooperator hub, for different values of $c_0$. The values of $r^{\textup{hub}}$ are computed by taking as hubs those vertices with degree equal or larger to the 97th and the 99th percentile of the degree distribution, respectively; they are shown as the top and bottom lines corresponding to the same value of $c_0$. $\xi_s=(\xi_2 + \xi_3)/2$, $\forall s$ (see main text for details). {\em Right}. The combined probability of Equation~(\ref{def_invasion}) that, in a $2$-uniform (solid lines) and $3$-uniform (dashed lines) DM hypergraph, an isolated defector $d$ survives and $w=1$ of its $\left|\Omega(d)\right|=8$ cooperator neighbors changes strategy, as a function of $\alpha$, for the values of $\left|\Omega\right|$ reported in the legend, $\kappa_d=1.75$, $\beta=1$ and $b=1$.}
  \label{inv_analysis}
\end{figure}

We focus on a cooperator hub with $m$-degree sequence $(k_{m_{\min}}^{hub},\dots,k_{m_{\max}}^{hub})$, surrounded by neighbors with average $k$-degree sequence $(\bar{k}_{m_{\min}},\dots,\bar{k}_{m_{\min}})$, where
\begin{equation}
  \bar k_m=\bar k_m(\{k_{m}^{hub}\})=\sum_{k_m}P(k_m\vert k_{m}^{hub})\,k_m
\end{equation}
Imposing the payoff of the hub playing as cooperator is equal to the average payoff of its neighbors playing as defectors, $f^{(hub)}=\bar{f}^{(d)}$, we get the threshold value $\alpha_{\textup{th}}^{\textup{hub}}$ above which, whenever the hub and any of its defector neighbors match for the strategy update, the survival of the hub is favored. Indicating with ${\bar{q}_0}^{(d)}=c_0N/(N-1)$ and $q_0^{(hub)}={\bar{q}_0}^{(d)}-1/(N-1)$ the initial ($t=0$) average fraction of cooperating neighbors, respectively, of the hub's defector neighbors and of the hub itself, from Equation~(\ref{payoff_dynamics}) we find
\begin{equation}
  \alpha_{\textup{th}}^{\textup{hub}} = \frac{{\ds\sum_{m=m_{\min}}^{m_{\max}}}k_m^{\textup{hub}}}{{\ds\sum_{m=m_{\min}}^{m_{\max}}}\left[\ds\frac{k_m^{\textup{hub}}}{m}+\frac{m-1}{m}\left(q_0^{(hub)}k_m^{\textup{hub}}-{\bar{q}_0}^{(d)}\bar k_m\right)\right]}
  \label{Hub_threshold}
\end{equation}
Considering $s$-uniform hypergraphs and putting $\bar k_s=\xi_s k_{s}^{hub}$, from Equation~(\ref{Hub_threshold}) follows
\begin{equation}
  r^{\textup{hub}} \equiv \frac{\alpha_{\textup{th}}^{\textup{hub}}}{\alpha_{\textup{cr}}} = \frac{1}{1+(s-1)(q_0^{(hub)}-{\bar{q}_0}^{(d)}\xi_s)}
\end{equation}
which is a monotone decreasing function of $s$ and, for $\xi_s<1-(N\,c_0)^{-1}$, of $c_0$ as well (left panel of Figure~\ref{inv_analysis}). Taking as hubs those vertices with degree equal or larger, respectively, to the 97th and the 99th percentile of the degree distribution, we get, for each value of $s$, a small interval in which, the values of $r(p)$ found in the simulations for $c_0=0.5$, fit well. In fact, we expect Equation~(\ref{Hub_threshold}) to work well for intermediate values of $c_0$, as commented in Appendix~\ref{sec:appendix_d}. Moreover, we expect the method to work for $\beta$ large enough, when the imitation process is more deterministic.

To evaluate Equation~(\ref{Hub_threshold}) we take $k_s^{\textup{hub}}$ as the average degree of those vertices classified as hubs and then compute the corresponding average value of $\xi_s$. Setting $c_0=0.5$, for HK hypergraphs we find $r^{\textup{hub}}$ between 0.71 and 0.79 for $s=2$, and between 0.55 and 0.65 for $s=3$; for DM hypergraphs we get $r^{\textup{hub}}$ between 0.72 and 0.79 for $s=2$, and between 0.56 and 0.65 for $s=3$.  In Figure~\ref{inv_analysis} (left panel), to visualize $r^{\textup{hub}}$ versus $s$, we reported the values taken by the former computed considering $\xi_s$ as independent of $s$ and fixed equal to $(\xi_2 + \xi_3)/2$, taking $\xi_2$ and $\xi_3$ as the average of the respective values found in the used hypergraphs. In fact, only $\xi_2$ and $\xi_3$ are accessible in HK and DM hypergraphs. Nevertheless, $\xi_s$ only decreases by about $0.01$ going from $s=2$ to $s=3$, therefore keeping it fixed is a sensible choice.

When $\alpha>\alpha_{\textup{th}}^{\textup{hub}}$, a cooperator hub is resilient to the invasion by any defector in its neighborhood and, by persisting in the cooperative state, gives rise to a cooperative community whose members sustain each other (for a similar dynamics on networks see \cite{santos}). In this way, this community is able to enlarge itself further or, at least, to resist to be invaded by the defectors in the neighborhoods, thanks to the local topological constraints. Remarkably, in HK and DM networks, by construction, the hubs are frequently very close to each other (first or second neighbors), making it easier for the cooperative community to thrive when more than one of them happens to be in the cooperative state. When, instead, the hub is a defector, its local proliferation is favored whatever is the value of $\alpha$. However, once a defector hub easily succeeded in converting all its lower-degree cooperator neighbors in defectors, none of the players in this defective community provide/receive any benefit playing within it. This prevents the defective community to cover the entire population, unless $\alpha$ is enough small, outside the critical region.

\unskip
\subsubsection{The Role of Isolated defectors}
\label{sec:isolated_defector}

Let us now consider the limit case in which a $\sigma_1$-strategist is totally surrounded ---up to its $z^{th}$ neighborhood, with $z$ large enough---  by $\sigma_2$-strategists, with $\sigma_1\neq \sigma_2$, to calculate the joint probability that the $\sigma_1$-strategist survives and some of the $\sigma_2$-strategists adopt strategy $\sigma_1$, as a function of $\alpha$. The probability $p_i(t)$ that a player of strategy $\sigma_i$ and neighbors $\Omega(i)$ changes its strategy, reads:
\begin{equation}
  p_i(t)=\frac{1}{\left|\Omega (i)\right|}\underset{\underset{\sigma_j\neq \sigma_i}{j\in \Omega (i)}}{\sum\;}\mathcal F\left(f^{(j)}(t)-f^{(i)}(t)\right)
  \label{p_i}
\end{equation}
In particular, considering a defector $d$ of degree pair $(k_2^{(d)},k_3^{(d)})$ as the isolated strategist, the probability for it to survive as a defector at time $t$ is equal to $1-p_d(t)$, in which the difference between the payoff $f^{(j)}(t)$ of a cooperator neighbor $j$ and its payoff $f^{(d)}(t)$ is given by:
\begin{equation}
  f^{(j)}(t)-f^{(d)}(t)=b\left[\alpha\left(k_2^{(j)}-\frac{1}{2} k_2^{(d)}+k_3^{(j)}-\frac{2}{3}k_3^{(d)}-a^{(j)}\right)-k_2^{(j)}-k_3^{(j)}\right]
  \label{payoffs_diff_isolated_def}
\end{equation}
where $a^{(j)}$ is equal to $1/2$ if $j$ is a neighbor through a link and to $\kappa^{(j)}_d/3$ if it is a neighbor through a triangle, where $\kappa^{(j)}_d$ is the number of triangles that $j$ shares with vertex $d$. Similarly, the probability that a cooperator neighbor $j$ becomes a defector is equal to $\mathcal{F}\left(f^{(d)}(t)-f^{(j)}(t)\right)/\left|\Omega (j)\right|$.

To get interesting insights it is sufficient to consider all the defector's neighbors (all cooperators) as equivalent to the average vertex in the neighborhood. Therefore, they have the same degree pair $(k_2^{(c)},k_3^{(c)})$, the same number of neighbors $\left|\Omega\right|$, the same $\kappa_d$ and, consequently, the same payoff $f^{(c)}(t)$; it makes sense because the neighbors are selected uniformly at random when updating the strategies. Then, the probability for $d$ of surviving as a defector and invading all at once $w\leq \left|\Omega (d)\right|$ neighboring sites reads
\begin{equation}
  (1-p_d(t))\cdot \left[\frac{1}{\left|\Omega\right|}\mathcal F\left(f^{(d)}(t)-f^{(c)}(t)\right)\right]^w
  \label{def_invasion}
\end{equation}

In Figure~\ref{inv_analysis} (right panel) we show this combined probability computed for $w=1$, $\left|\Omega(d)\right|=8$, different values of $\left|\Omega\right|$, $\kappa_d=1.75$, and in the presence of either only links ($p=0$) or only triangles ($p=1$). These values correspond to neighborhoods compatible with DM and HK hypergraphs (see Appendix~\ref{sec:appendix_e} for details). Interestingly, the defector should prefer interacting through triangles whenever $\left|\Omega(d)\right|>\left|\Omega\right|$ and through links otherwise. This is simply understood observing that, thanks to the percolation of 3-cliques (or triangles) present in HK and DM networks (hypergraphs), ${\left.k_3^{(d)}\right|}_{p=1}$ is close to ${\left.k_2^{(d)}\right|}_{p=0}$; in particular, ${\left.k_3^{(i)}\right|}_{p=1}={\left.k_2^{(i)}\right|}_{p=0}-1$, $\forall i$, for DM structures. Since when $d$ plays within a triangle obtains two times what it gets playing within a link, it is advantageous playing through triangles whenever it possesses a sufficient number of neighbors. In Figure~\ref{inv_analysis} we also note that, for $\left|\Omega\right|$ sufficiently smaller than $\left|\Omega(d)\right|$, the probability of proliferating for the defective strategy counter-intuitively slightly increases with $\alpha$ (or $r$), even above the mean-field critical point ($r=1$).

Taken together, these observations suggest that such dependence on $\alpha$ may contribute substantially to the reported enlargement of the critical region when $p$ is increased. Specifically, on one hand, the increasing of $\alpha$ always brings the cooperators' performance closer to the defectors' in any local instance of the game, making it easier for cooperators to invade (with certainty, eventually); on the other hand, defectors with high degrees are increasingly able to invade their neighborhoods, containing the diffusion of the cooperative strategy. This competitive dynamics, existing only for heterogeneous structures, is thus expected to delay the point at which cooperators succeed in fully invading the population, stretching out the transition, which is almost abrupt in homogeneous random hypergraphs.

Next, we will explain why this competition gets fiercer when $p$ is increased, enlarging even more the transition region.
From Equation~(\ref{payoffs_diff_isolated_def}) we can calculate the condition that the degrees of the defector and of its cooperator neighbors have to fulfill in order that $f^{(c)}-f^{(d)}$ decreases when $\alpha$ increases, thus raising the competition. Taking, as done before, all the defector's neighbors (all cooperators) as equivalent, by imposing $\partial(f^{(c)}-f^{(d)})/\partial\alpha<0$, we get
\begin{equation}
  \frac{\left|\Omega\left(d\right)\right|+1}{\left|\Omega\left(d\right)\right|}\left(\frac{1}{2} k_2^{(d)}+\frac{2}{3} k_3^{(d)}\right)>k_2^{(c)}+k_3^{(c)}
\label{general_condition}
\end{equation}
where $|\Omega (d)| = k_2^{(d)}+2k_3^{(d)}/{\kappa_d}$ is the number of neighbors of the isolated defector, and $\kappa_d$ is the number of triangles that each of the cooperator neighbors shares with it. Since $(|\Omega (d)|+1)/|\Omega (d)|\leq3/2$, Equation~(\ref{general_condition}) requires some overall degree heterogeneity in favor of the isolated defector. Instead, as anticipated, for a regular hypergraph it always holds $\partial(f^{(c)}-f^{(d)})/\partial\alpha>0$. Thus, when in a heterogeneous structure a defector possesses enough higher degrees with respect to the surrounding cooperators, its survival is favored when $\alpha$ is increased, smearing the transition.

To get some understanding about how this phenomenon depends on $p$, i.e., on the order of the interactions, we refer to hypergraphs that become 3-uniform when $p=1$, and 2-uniform when $p=0$, like the ones we have used. Then, Equation~(\ref{general_condition}) splits into the following conditions for $p=0$ and $p=1$, respectively:
\begin{align}
  \frac{k_2^{(d)}}{k_2^{(c)}} &> 2-\frac{1}{k_2^{(c)}}
  \label{splitted_conditiona}
  \\
  \frac{k_3^{(d)}}{k_3^{(c)}} &> \frac{3}{2}-\frac{\kappa_d}{2k_3^{(c)}}
  \label{splitted_conditionb}
\end{align}

Equating the expressions taken by $|\Omega (\sigma)|$ for $p=0$ and $p=1$, the identity ${\left.k_3^{(\sigma)}\right|}_{p=1}={\left.k_2^{(\sigma)}\right|}_{p=0}\cdot \kappa_\sigma/2$, to be read as $k_3^{(\sigma)}$ computed when $p=1$, given $k_2^{(\sigma)}$ when $p=0$, follows. We can then rewrite Equation~(\ref{splitted_conditionb}) in terms of $k_2^{(d)}$ and $k_2^{(c)}$ and directly compare it with Equation~(\ref{splitted_conditiona}). Equation~(\ref{splitted_conditionb}) becomes
\begin{equation}
  \frac{k_2^{(d)}}{k_2^{(c)}}>\frac{3}{2}\frac{\kappa_c}{\kappa_d}-\frac{1}{k_2^{(c)}}
  \label{condition_b}
\end{equation}
where the degrees are calculated for $p=0$. It follows that Equation~(\ref{condition_b}) is a weaker condition than Equation~(\ref{splitted_conditiona}), i.e., Equation~(\ref{splitted_conditiona}) $\Rightarrow$ Equation~(\ref{condition_b}, if and only if $\kappa_d/\kappa_c>3/4$. Remarkably, whenever the latter inequality is fulfilled, $\partial(f^{(c)}-f^{(d)})/\partial\alpha<0$ is more negative for $p=1$ than for $p=0$. Expressing $\kappa_\sigma$ in terms of the local clustering coefficient $C(\sigma)$ and the link-degree $k_2^{(\sigma)}$ as $\kappa_\sigma=[C(\sigma)\cdot(k_2^{(\sigma)}-1)]_{p=0}$ (see Appendix~\ref{sec:appendix_e} for the derivation), we finally get
\begin{equation}
  \mbox{Equation}~(\ref{splitted_conditiona})\Rightarrow \mbox{Equation}~(\ref{condition_b}) \;\;\;\ \Longleftrightarrow \;\;\;\ \left[\frac{C(d)\cdot(k_2^{(d)}-1)}{C(c)\cdot(k_2^{(c)}-1)}> \frac{3}{4}\right]_{p=0}
  \label{implication}
\end{equation}

According to the inequalities in Equations~(\ref{splitted_conditiona}) and~(\ref{condition_b}), the most interesting case is obtained when $k_2^{(d)}>k_2^{(c)}$. Then, Equation~(\ref{implication}) can be satisfied even when the isolated defector possesses a smaller clustering coefficient than its cooperators neighbors. Specifically, HK and DM models grow networks whose local clustering coefficient $C$ scales with the 2-degree $k_2$ as $C(k_2)\sim k_2^{-\gamma}$, where $\gamma$ is around 0.8 for HK ($P_t=1$) and exactly 1 for DM (see Appendix~\ref{sec:appendix_e}). Substituting in Equation~(\ref{implication}), one finds that the above implication holds whenever $k_2^{(d)}>k_2^{(c)}$, given $\gamma\leq 1$.

If we now ask for $k_2^{(d)}$ such that, for a given $k_2^{(c)}$, Equation~(\ref{condition_b}) is satisfied but Equation~(\ref{splitted_conditiona}) is not, we find the condition
\begin{equation}
  \frac{3}{2}\frac{\kappa_c}{\kappa_d}k_2^{(c)}-1<k_2^{(d)}\leq 2k_2^{(c)}-1
  \label{inequalities}
\end{equation}
where the degrees are calculated for $p=0$. Therefore, if $k_2^{(d)}$ takes an intermediate value in the sense of Equation~(\ref{inequalities}), with respect to $k_2^{(c)}$, Equation~(\ref{condition_b}) is satisfied but Equation~(\ref{splitted_conditiona}) is not, allowing the competitive dynamics to take place for $p=1$ but not for $p=0$. As shown in Appendix~\ref{sec:appendix_e}, such intermediate neighborhoods actually exist in the structures we made use of. It must be noted that, for $k_2^{(d)} > 2k_2^{(c)}-1$, the competitive dynamics exists for $p=0$ as well. However, as already observed, due to the linear dependence of $f^{(c)}-f^{(d)}$ on $\alpha$, $\partial(f^{(c)}-f^{(d)})/\partial\alpha<0$ is more negative for $p=1$ than for $p=0$ whenever $k_2^{(d)} > k_2^{(c)}$ (corresponding to $\kappa_d/\kappa_c>3/4$), making the competitive dynamics always fiercer in the former case.

Ultimately, this analysis shows that in heterogeneous and clustered structures, like the ones we used here, the competitive dynamics takes place more frequently and intensively in the presence of 3-way ($p>0$) than 2-way ($p=0$) interactions, in accordance with the observed enlargement of the transition region when increasing $p$. It must be said that our analysis does not exclude the existence of other mechanisms that could further contribute to the observed enlargement.

\unskip
\subsection{Heterogeneous Populations: The Role of Topological Traps}
\label{sec:traps}

\begin{figure}[tb!]
  \centering
  \includegraphics[width=0.85\linewidth]{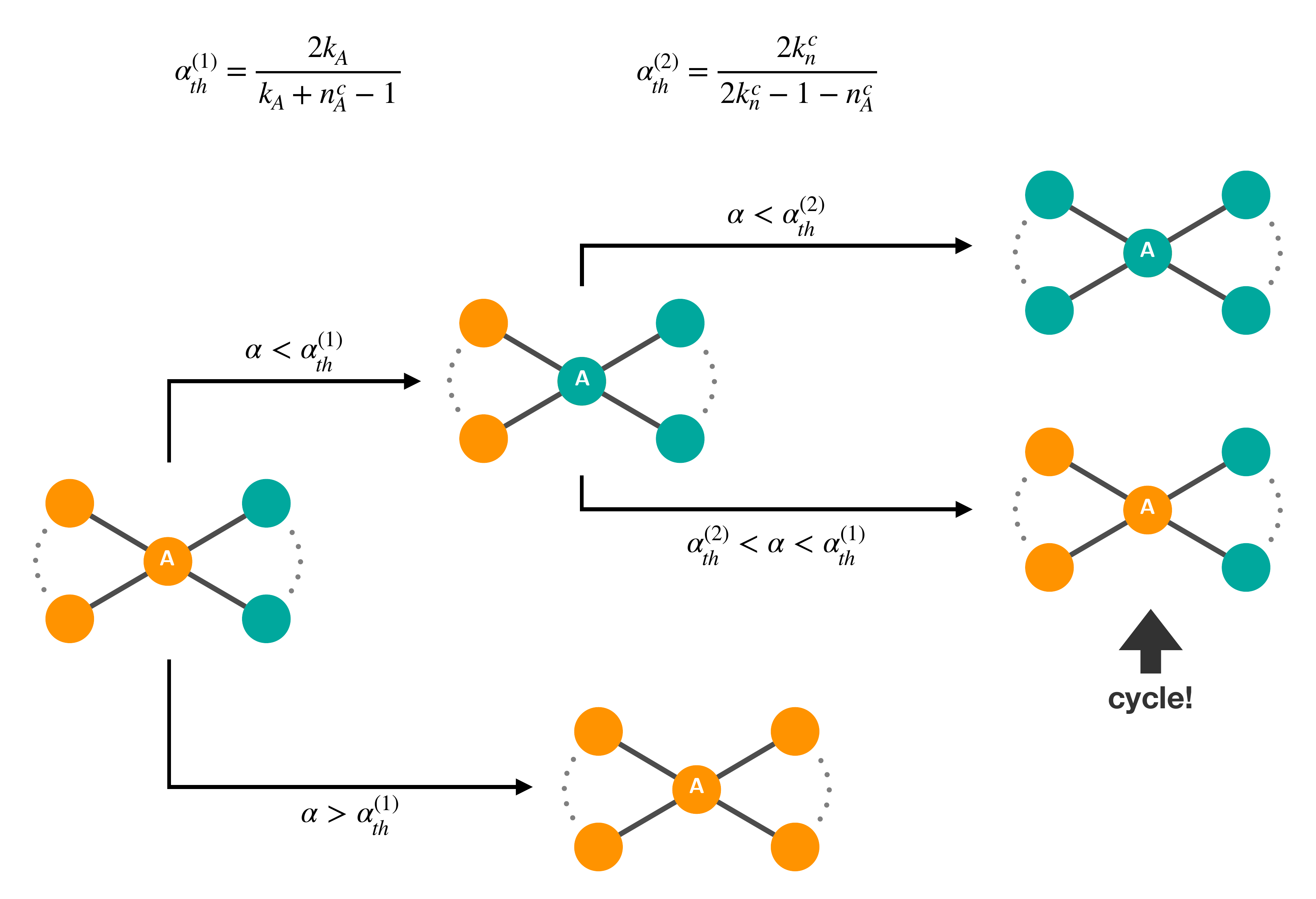}
  \caption{Generic bottleneck trap at work in presence of only links. Cooperators are shown in orange. Vertex $A$ is the bridge vertex, $k_A$ is its degree and $n_A^c$ is its number of cooperator neighbors (all considered with same degree $k_n^c$ to get a unique $\alpha_{\textup{th}}^{\left(2\right)}$ and make clearer the illustration). For $\alpha_{\textup{th}}^{\left(2\right)}<\alpha<\alpha_{\textup{th}}^{\left(1\right)}$, after some time steps, the configuration is likely to return to the initial state, giving rise to an endless cycle that avoids both strategies to spread through the trap. Note that a minimal degree heterogeneity among vertex $A$ and its neighbors must exists to make the trap work.}
  \label{trap_example}
\end{figure}

When running a dynamical process on top of a connected structure, there could be microscopic and mesoscopic configurations which operate as \emph{topological traps}, blocking the system in a certain state. Traps represent here the only mechanism able to give rise to asymptotic heterogeneous populations in which cooperators and defectors co-exist. A trap can have any size and complexity, i.e., it can involve any number of vertices and develop over many time steps. A particularly simple class of traps is the bottleneck-type's, consisting of a sub-hypergraph in which some vertices act as bridges between groups of vertices of opposite strategy. In rank-3 hypergraphs, they consist of some bridge-vertices gluing links and/or triangles, as shown in Figure~\ref{trap_example}.

A trap comes into play when the strategies are updated. According to our setup, it is only the 2-section of the hypergraph that matters for the strategy update, and the presence of higher-order interactions solely affects the ranges of values of $\alpha$ for which a local structure can act as a trap. Then, the effect of a bottleneck trap is fully specified by the comparison between the chosen value of $\alpha$ and the threshold values $\alpha_{\textup{th}}$s, defined as those values at which a cooperator and a defector, each with its $m$-degree sequence, get the same final payoff. Any trap other than bottleneck-type is specified simultaneously by a set of coupled $\alpha_{\textup{th}}$s, one for any link of the 2-section contributing to the trap.

Whatever is the complexity of a trap, its effectiveness is strictly related to the value of the parameter $\beta$. A trap is able to act if the selection is strong enough, i.e., for high enough values of $\beta$. Indeed, $\beta$ weighs the proportional contribution of the difference of the payoffs to the update rule. The smaller is $\beta$, the more random the strategy update is, being increasingly disentangled from the payoffs resulting from the interactions. Accordingly, the MC simulations made for $\beta=0.1$ provide convergence times notably shorter than the ones found for $\beta=1$. Furthermore, we see also a slight reduction of structural reciprocity when lowering $\beta$, meaning that the structures develop some traps able to further anticipate the survival of cooperation towards slightly lower values of $r$.

A critical value of $\alpha$ can either satisfy or not some of the local conditions for producing a trap. In other words, the found effective densities of cooperators can be the result of an average of both homogeneous final states (the evolutionary stable states) and/or heterogeneous ones (produced by the traps). It was already unveiled in \cite{roca2} that, for the entire set of 2$\times$2 social dilemmas played on different real networks, the presence of bottleneck traps is strictly related to the degree heterogeneity and the lack of redundant paths.

\begin{figure}[tb!]
  \centering
  \includegraphics[width=.46\linewidth]{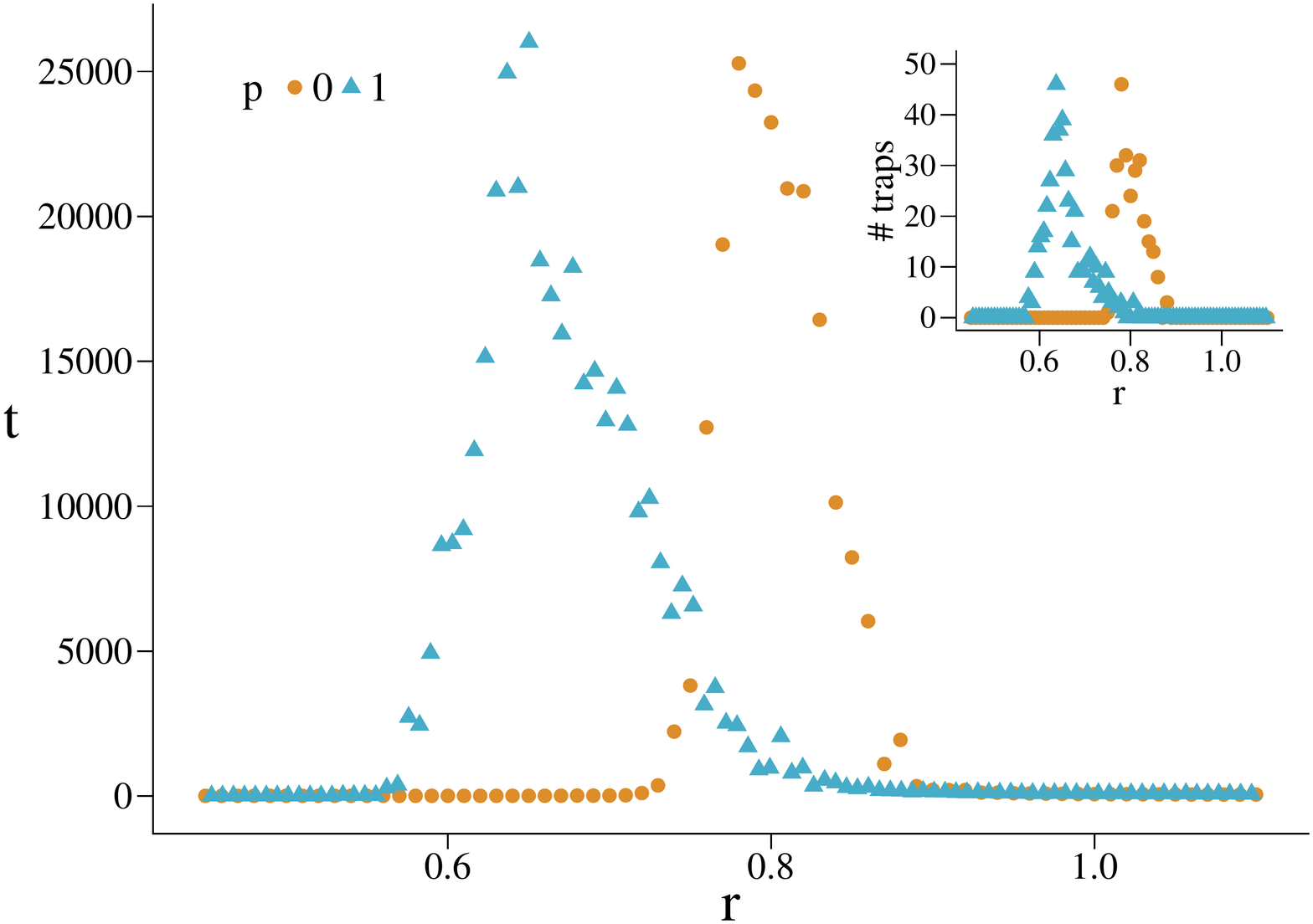}
  \hspace{0.5cm}
  \includegraphics[width=.46\linewidth]{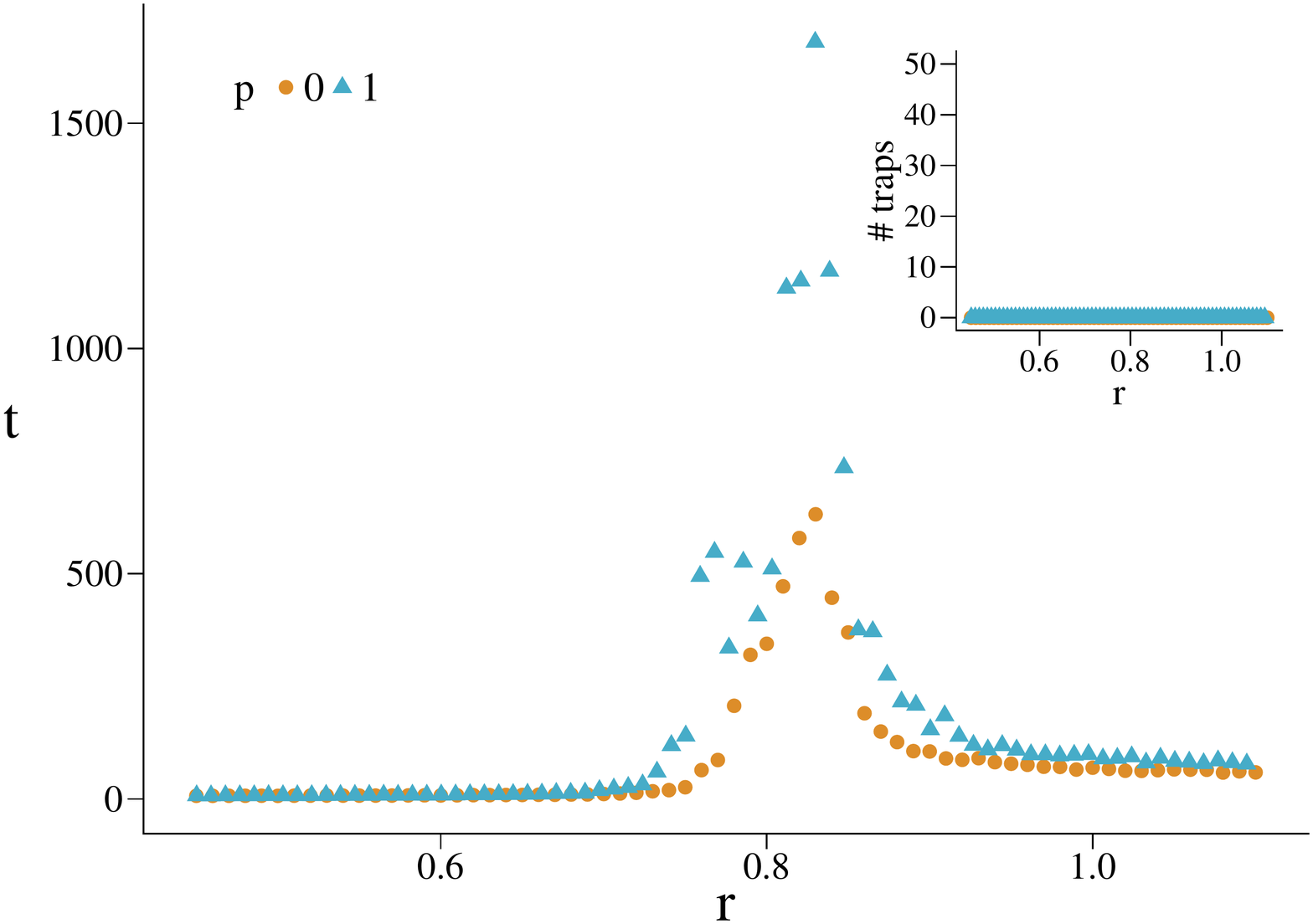}
  \caption{Convergence time versus $r$ for the hypergraphs generated from two differently rewired versions of the used DM networks, with $t_{\max}=5\times10^4$. The inset plots report the number of times the system has not converged (i.e., the number of times the system got trapped). {\em Left}. Randomization preserving degree and local clustering: traps are frequent in the critical regions. {\em Right}. Randomization preserving second-order degree correlations: no traps left. The values of $p$ are specified in the legend; $c_0=0.5$, $\beta=1$. Note that, in DM hypergraphs, bottleneck traps cannot exist, since they boost a complete percolation of triangles.}
  \label{rewiring}
\end{figure}

To characterize the generic traps for the PGG played on rank-3 simple hypergraphs, similarly to what is done in \cite{roca2}, we perform two different randomization procedures of the networks (see Section~\ref{sec:methods} for details). One randomization preserves both the degree and the local clustering coefficient of each vertex, while the other preserves the degree correlations up to the second order, i.e., up to the joint $3K$-distribution $P(k,k',k'')$. Clustering coefficient and second-order degree correlations are clearly related, but preserving one does not fully imply preserving the other automatically. Now, since the action of a trap is defined in multiple time steps (two at least) in which are involved several vertices with overlapping neighborhoods, the degree correlations of order higher than two should be important in the trapping process. Destroying those correlations we expect the traps to dissolve. Quite surprisingly, we find that what really matters for the presence of traps other than the bottleneck-type, is the high value of the local clustering coefficient (see Figure~\ref{rewiring}). Preserving the $P(k,k',k'')$ distribution without caring of the local clustering, unravels the initially present traps. We find this for both DM and HK hypergraphs. On the contrary, the presence of bottleneck traps is clearly negatively correlated with clustering; they are found to be dominant for lowly clustered networks (scarce of redundant paths) exhibiting strong degree heterogeneity, in line with \cite{roca2}.

\section{Conclusions}
\label{sec:conclusions}

Collective and, in particular, cooperative phenomena, in which many individuals take part as a group, are observed in many real systems, among which human society stands out. In this work we investigated systematically the effect that group (higher-order) interactions have on the evolution of cooperation in structured populations. We studied the dynamics arising from a population of individuals interacting by playing a public goods game on top of hypergraphs. While the theory we developed is general, the numerical simulations specifically regarded rank-3 hypergraphs. In this way, we distinguish whether an element, when a member of a group (clique), interacts separately, in pair, with each of the other members, or interacts simultaneously with all of them.

To get the hypergraphs, we made use of an algorithm preserving the pairwise projection of the structure, which is crucial to satisfy the constraints implicit in the network topology. In fact, although a pairwise projection can correctly describe who interacts with whom, it cannot account for different types of interactions, the temporal order in which the latter occur, and whether interactions involve pairs or larger groups of units. Additionally, the 2-section imposes the maximum possible rank one could reach by generalizing the herein given procedure. The existence of a maximal size for groups is usually related to a saturation restriction due to some unbalanced cost in forming larger groups. This generally depends on the nature of the considered system, and addressing its form goes beyond the scope of this work. Here, we just recognize the existence of such a restriction as encoded in the 2-sections of the used hypergraphs, and preserve it with our generative method.

Monte Carlo simulations clearly indicated that heterogeneous structures, like the ones obtained from HK and DM networks, are able to sustain cooperation for notable ranges of values of the synergy parameter below its critical value. This holds for unstructured populations, whatever the conversion fraction, related to the order of the interactions. Our findings, therefore, extend the phenomenon of network reciprocity to rank-3 hypergraphs.

Remarkably, we show a clear evidence that the sustenance of cooperation is stronger when the fraction of 3-way interactions is increased. Indeed, we attained a monotonous decrease of structural reciprocity with the increasing of conversion fraction, for hypergraphs derived from both HK and DM networks. Morever, the level of cooperation is always larger for the hypergraph. It is key noting that this enhanced reciprocity is exclusively due to the substitution of some closed triads of first-order interactions (3-cliques) with unique second-order interactions (triangles). In other words, given a heterogeneously-structured population, cooperation thrives more easily if individuals interacting separately, in pairs, within a closed triad are made to interact all together. It should be noticed that, considering the synergy factor as independent of the group sizes, we did not account for explicit synergistic or anti-synergistic effects that could increase or decrease the found levels of reciprocity. The improvement found here, rather, can be referred to as a topological synergistic effect allowed by the evolutionary dynamics.

As could be expected from what is known on networks, an important requirement for a significant reciprocity is the heterogeneity of connections; a separate analysis, not addressed in this work, would be needed for higher-order generalizations of spatial networks. To prove this, we performed an invasion analysis able to provide a good estimate for the critical point of emergence of cooperation, for both hypergraphs with rank equal to~2 and to~3. More generally, the analysis indicated that, the higher the order of the interactions, the stronger the structural reciprocity is. To be more precise, the marked enhancement of reciprocity has been found for heterogeneous structures with a null or slightly negative assortativity, in line with what is reported for networks. According to our analysis, we expect a weaker beneficial effect on hypergraphs derived from assortative networks.

Furthermore, the invasion analysis allowed us to justify, in comparison with the almost abrupt transition found in homogeneous structures, the enlargement of the transition region observed for the used heterogeneous hypergraphs. Relying on their locally-clustered structure, it also allowed to partially explain why the enlargement grows with the increasing of conversion fraction.

Regarding the heterogeneity of strategies in the asymptotic population, we found the presence of topological traps preventing the system to converge to a uniform state. We characterized those traps by means of two randomization procedures. In the used clustered structures, we attended the break-up of the present traps whenever the chosen randomization did not preserve the local clustering, though still preserving second-order degree correlations. We show that traps represent the only way to get a non-uniform asymptotic population.

Finally, by tuning the selection pressure, we found an improvement towards cooperation in conditions of stronger selection. This is precisely what we expect from how topological traps work. Remarkably, this shows up the contribution coming from some traps to enlarge the region where cooperative behaviors can survive.

\section{Methods}
\label{sec:methods}

\unskip
\subsection{Generating Rank-3 Simple Hypergraphs}
\label{sec:algorithm}

\begin{figure}[tb!]
  \centering
  \includegraphics[width=0.85\linewidth]{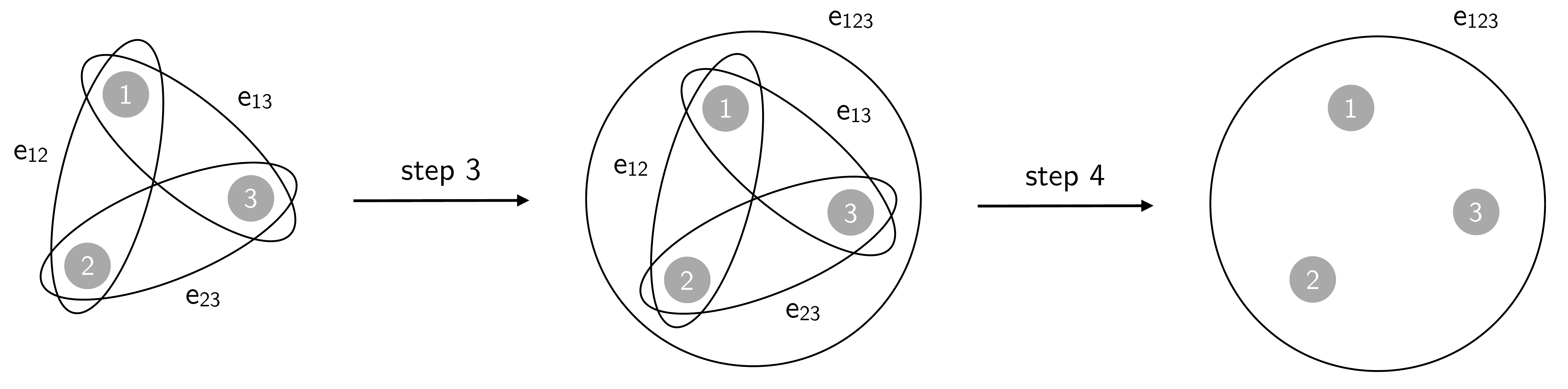}
  \caption{Illustration of a single instance of steps 3 and 4 of the algorithm used to generate rank-3 simple hypergraphs starting from graphs. Note that, removing step~4, we would end up with a simplicial complex.}
  \label{gen_hyp}
\end{figure}

The algorithm to generate rank-3 simple hypergraphs proceeds as follows:

\begin{enumerate}[leftmargin=*,labelsep=4.9mm]
  \item A simple network $g=\left(V_0,E_0\right)$ is generated through some model;
  \item From the set of all the 3-cliques in it, a fraction $0\leq p\leq 1$ of them is picked uniformly at random;
  \item To each of the picked 3-cliques a triangle is associated: if $e_{12}=\left\{v_1,v_2\right\},e_{23}=\left\{v_2,v_3\right\},e_{13}=\left\{v_1,v_3\right\}\subset E_0$ are the three edges forming a chosen 3-clique over the subset $\left\{v_1,v_2,v_3\right\}\subset V_0$ of vertices, then the hyperedge (triangle) $e_{123}=\left\{v_1,v_2,v_3\right\}$ is added to $E_0$;
  \item To obtain a simple hypergraph, after a fraction $p$ of 3-cliques has been converted to triangles, an edge is removed if it is subset of at least one triangle: the three edges $e_{12},e_{23},e_{13}$, being subsets of $e_{123}$, are removed from $E_0$.
\end{enumerate}
This procedure generates a rank-3 simple hypergraph $H=\left(V,E\right)$, where $V=V_0$ and $E$ is the new set of hyperedges constructed from $E_0$ through the steps 2--4. The initial network $g$ is the 2-section of $H$. See Figure~\ref{gen_hyp} for an illustration of the procedure.

Since the algorithm relies on the amount of 3-cliques present in the base network $g$, the more clustered is the latter, the stronger will be the effect of increasing the value of $p$. In HK ($P_t=1$) and DM networks, every link is part of at least a 3-clique. Therefore, by tuning $p$, they allow us to explore the entire range of hypergraphs from the 2-uniform ($p=0$) to the 3-uniform ($p=1$).

\unskip
\subsection{Monte Carlo Simulations}
\label{sec:mc_simu}
Every simulation starts by picking, uniformly at random, a fraction $c_0\equiv c(t=0)$ of cooperators. For each fixed set of parameters $\beta$, $c_0$ and $p$, we run 100~simulations over a generated hypergraph and take the average of both the asymptotic cooperators density and the convergence time, i.e., the number of time steps the process took to stop. We do it for~10 realizations of the same kind of hypergraph (i.e., stemming from the same network model), and then average again over them. Whenever the maximum number of time steps (set to $2\times10^4$) is reached without converging to one of the absorbing states $c=0$ or $c=1$, the last value of $c$ is considered.

\unskip
\subsection{Randomization Procedures}
\label{sec:randomization}

We briefly summarize the two randomization procedures implemented for the characterization of topological traps. The selected number of randomizing steps was $10^5$ for each network.

\unskip
\subsubsection{Preserving Local Clustering Coefficient}
\label{sec:randomization_1}

Through the following procedure one can randomize a network preserving, besides the degree, also the local clustering coefficient of each vertex. At each step, two vertices are uniformly chosen at random and their local clustering coefficient is computed. Two links, one for each vertex, are randomly selected and interchanged. If the clustering coefficients of the two vertices remain unchanged, the randomizing step is accepted; otherwise, another two vertices are drawn and the randomization is tried again.

\unskip
\subsubsection{Preserving Second-Order Degree Correlations}
\label{sec:randomization_2}

This procedure allows to obtain a randomized network with exactly the same degree correlations up to the second-order, i.e., preserving the joint $3K$-distribution $P(k,k',k'')$. At each step, two vertices are uniformly chosen at random and the pair is accepted if only the vertices have equal degree. Then, two links, one for each vertex, are peaked at random (an interchange performed now would ensure the conservation of the $2K$-distribution $P(k,k')$). The degrees of the two vertices at the end of the two selected links are compared and, if they match, the interchange is performed (now conserving also $P(k,k',k'')$); elseways, two new initial vertices are chosen and the randomization is tried again.

\vspace{6pt}


\authorcontributions{Conceptualization, A.A.; methodology, A.A., S.G., J.M. and G.B.; investigation, G.B.; writing--original draft preparation, G.B.; supervision, A.A, J.M. and S.G. All authors have read and agreed to the published version of the manuscript.}

\funding{We acknowledge support by Ministerio de Econom\'ia y Competitividad (grants PGC2018-094754-B- C21 and FIS2015-71929-REDT), Generalitat de Catalunya (grant 2017SGR-896), and Universitat Rovira i Virgili (grant  2019PFR-URV-B2-41). A.A.\ acknowledges also ICREA Academia and the James S.\ McDonnell Foundation (grant 220020325).}

\conflictsofinterest{The authors declare no conflict of interest.}


\appendixtitles{yes} 
\appendix
\section{Critical Point for Uniform Complete Simple Hypergraphs}
\label{sec:appendix_a}

It is possible to find analytically the critical value of $\alpha_{\textup{cr}}^m$ at which $c_\infty=0.5$ for a finite $m$-uniform complete simple hypergraph with $N$ vertices, in the limit of strong selection ($\beta\,b\gg 1$). The previous condition makes deterministic the update of the strategies whenever two players with different strategies match. For large values of $N$ the transition becomes discontinuous and $\alpha_{\textup{surv}}^m$, above which $c_\infty>0$, converges to $\alpha_{\textup{cr}}^m$. The following proposition holds:

\begin{Proposition}
Given a complete $m$-uniform simple hypergraph ($m\geq 2$) with $N$ vertices, the critical value $\alpha_{\textup{cr}}^m$, in the limit $\beta\,b\gg 1$, is given by
\begin{equation}
  \alpha_{\textup{cr}}^m = m\frac{N-1}{N-m}
  \label{proposition1}
\end{equation}
where $\beta$ is the noise parameter of the Fermi distribution function and $b$ is the fixed contribution a cooperator puts at each time step.
\end{Proposition}

\begin{proof}[Proof of Proposition A1]
The expression for $\alpha_{\textup{cr}}^m$ follows imposing that the probabilities $p_{d\rightarrow c}$ and $p_{c\rightarrow d}$ that, respectively, a defector becomes a cooperator and vice versa, coincide.  Being the hypergraph complete, each vertex possesses $\binom{N-1}{m-1}$ neighbors. Therefore, the payoff $f^{(\sigma)}$ got by any player with strategy $\sigma$ reads
\begin{equation}
  f^{(\sigma)} = b\binom{N-1}{m-1}\left[\left(\frac{\alpha}{m}-1\right)\sigma+\alpha\frac{m-1}{m}q^{(\sigma)}\right]
\end{equation}
where $q^{(\sigma)}=q^{(c)}=c_0\,N/(N-1)-1/(N-1)$ for a cooperator and $q^{(\sigma)}=q^{(d)}=c_0\,N/(N-1)$ for a defector. Defining $\Delta f\equiv f^{(c)}-f^{(d)}$, we get
\begin{align}
  p_{d\rightarrow c} &= q^{(d)}\mathcal F\left(\Delta f\right) = c_0\frac{N}{N-1}\mathcal F\left(\Delta f\right)
  \\
  p_{c\rightarrow d} &= \left(1-q^{(c)}\right)\left(1-\mathcal F\left(\Delta f\right)\right) = \left(1-c_0\frac{N}{N-1}+\frac{1}{N-1}\right)\left(1-\mathcal F\left(\Delta f\right)\right)
\end{align}
where $\mathcal F$ indicates the Fermi distribution function. By equating $p_{d\rightarrow c}$ and $p_{c\rightarrow d}$, and solving with respect to $\alpha$, we get
\begin{equation}
\alpha_{\textup{cr}}^m = m\,\frac{N-1}{N-m}\,\frac{\ds\binom{N-1}{m-1}-\frac{1}{\beta b}\ln\left(\ds\frac{c_0}{1-c_0}\right)}{\ds\binom{N-1}{m-1}}
\label{proposition0}
\end{equation}
Taking $\beta\,b\gg 1$, Equation~(\ref{proposition1}) follows.

Note that, only for $c_0=0.5$, Equation~(\ref{proposition0}) automatically reduces to Equation~(\ref{proposition1}), meaning that the prediction is exact for any value of $\beta\,b$ in this case. Accordingly, the prediction is very precise also for low values of $\beta$ whenever $c_0$ does not differ much from 0.5.
\end{proof}

In the limit $N\rightarrow\infty$, from Equation~(\ref{proposition1}), we recover the critical value $\alpha_{\textup{cr}}^m = m$, which holds for an infinite well-mixed population, in accordance to Equation~(\ref{MF_critical_point}) evaluated for a $m$-uniform hypergraph.

\section{Simulations for Further Values of the Initial Fraction of Cooperators}
\label{sec:appendix_b}

\begin{figure}[tb!]
  \centering
  \includegraphics[width=.46\linewidth]{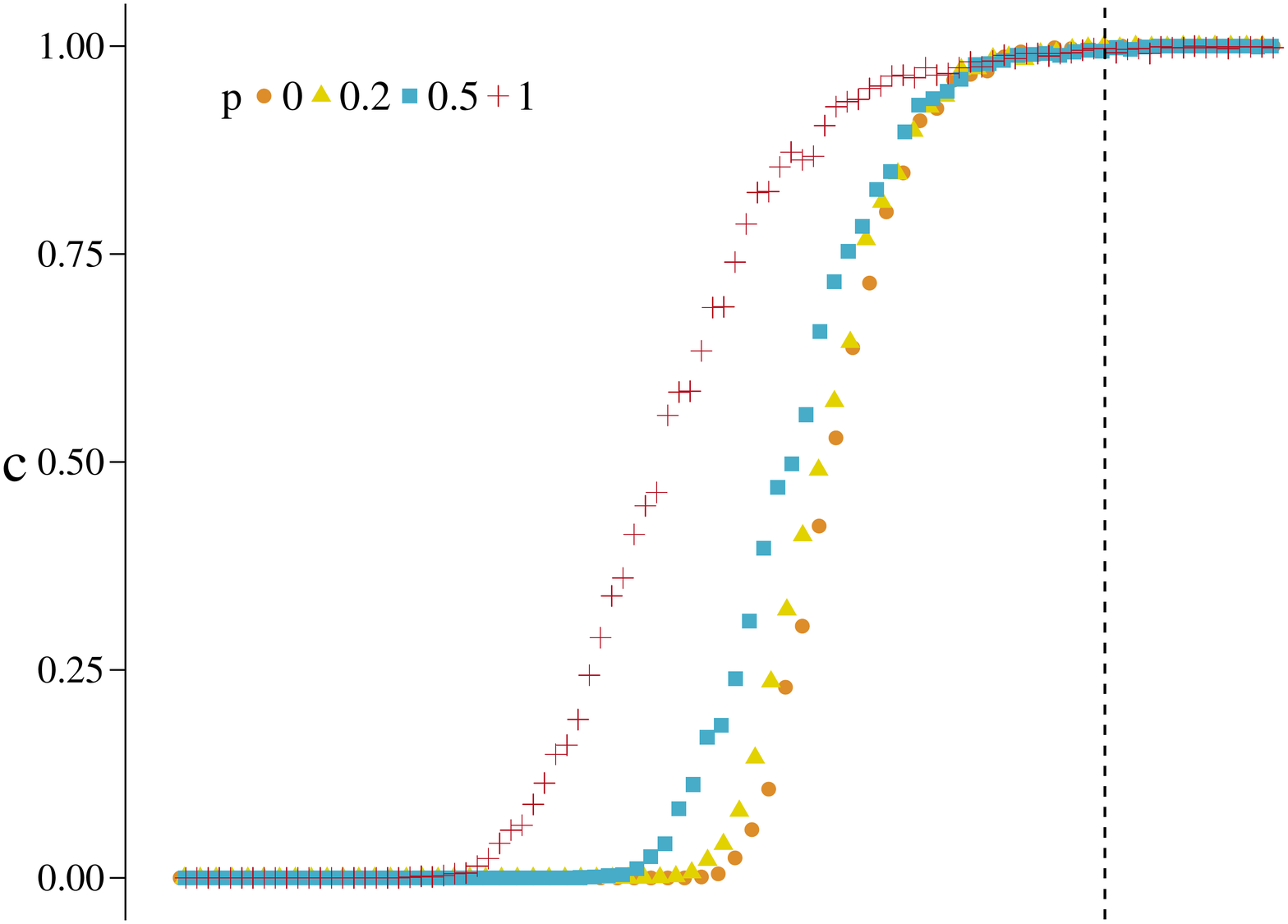}
  \hspace{0.5cm}
  \includegraphics[width=.46\linewidth]{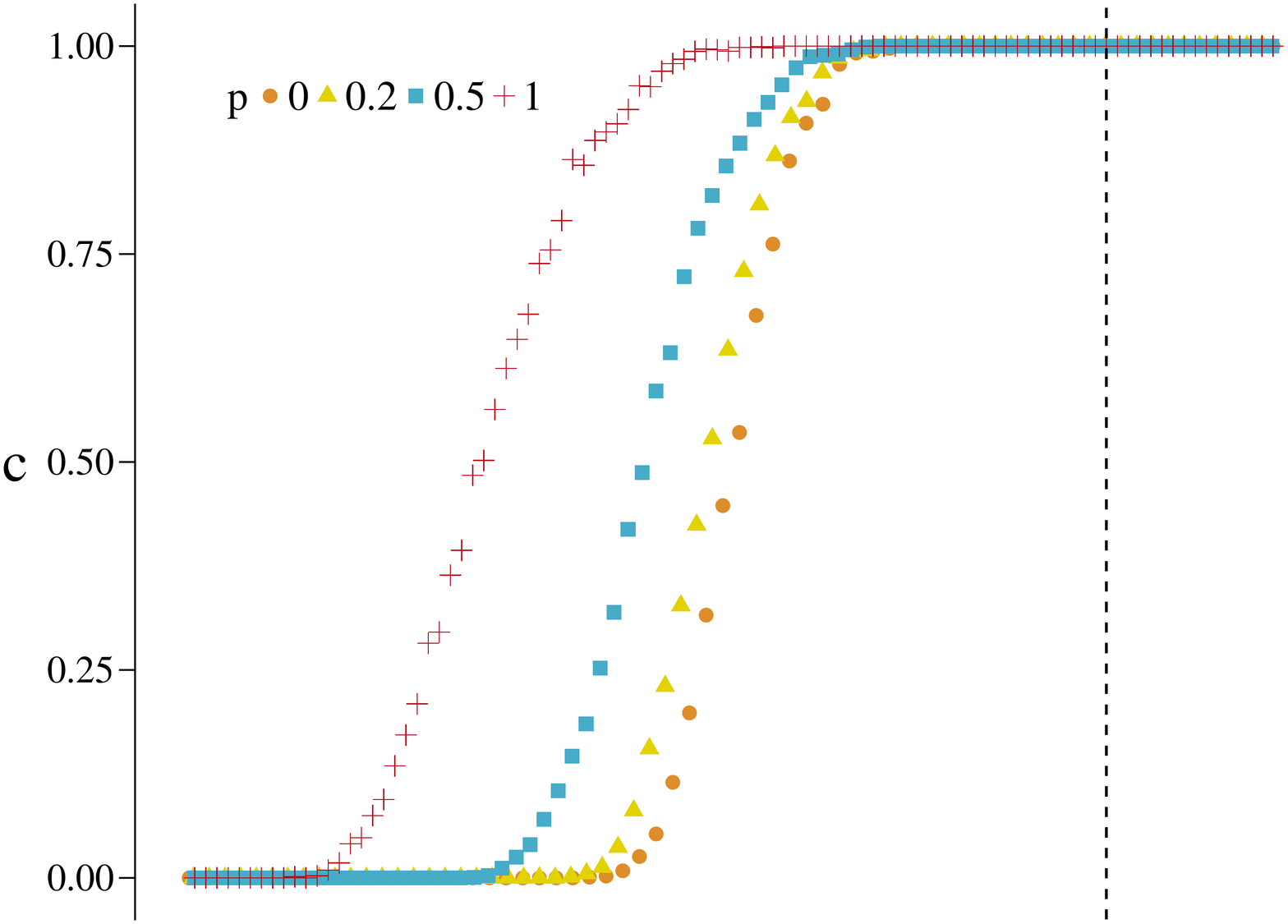}
  \\
  \includegraphics[width=.46\linewidth]{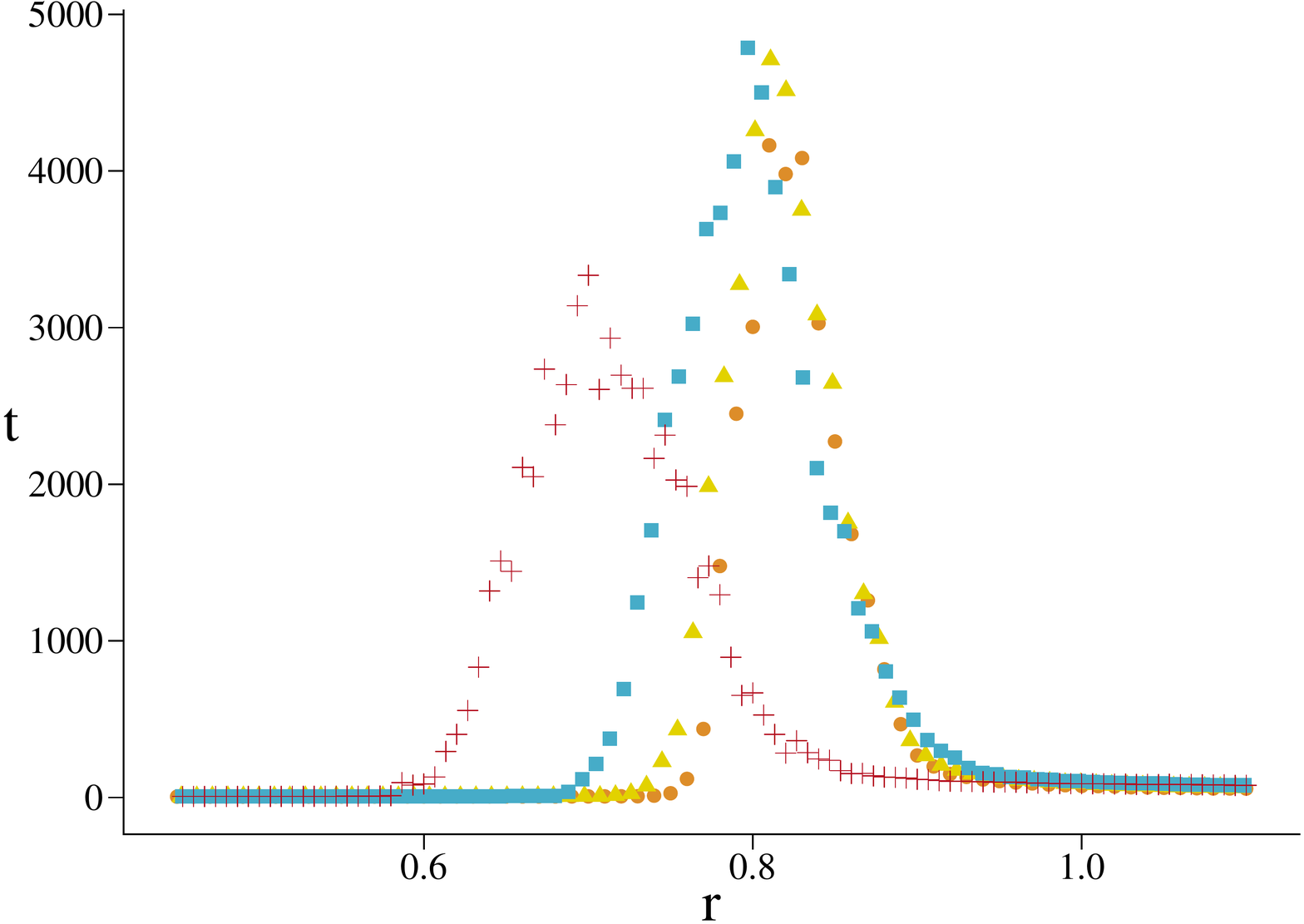}
  \hspace{0.5cm}
  \includegraphics[width=.46\linewidth]{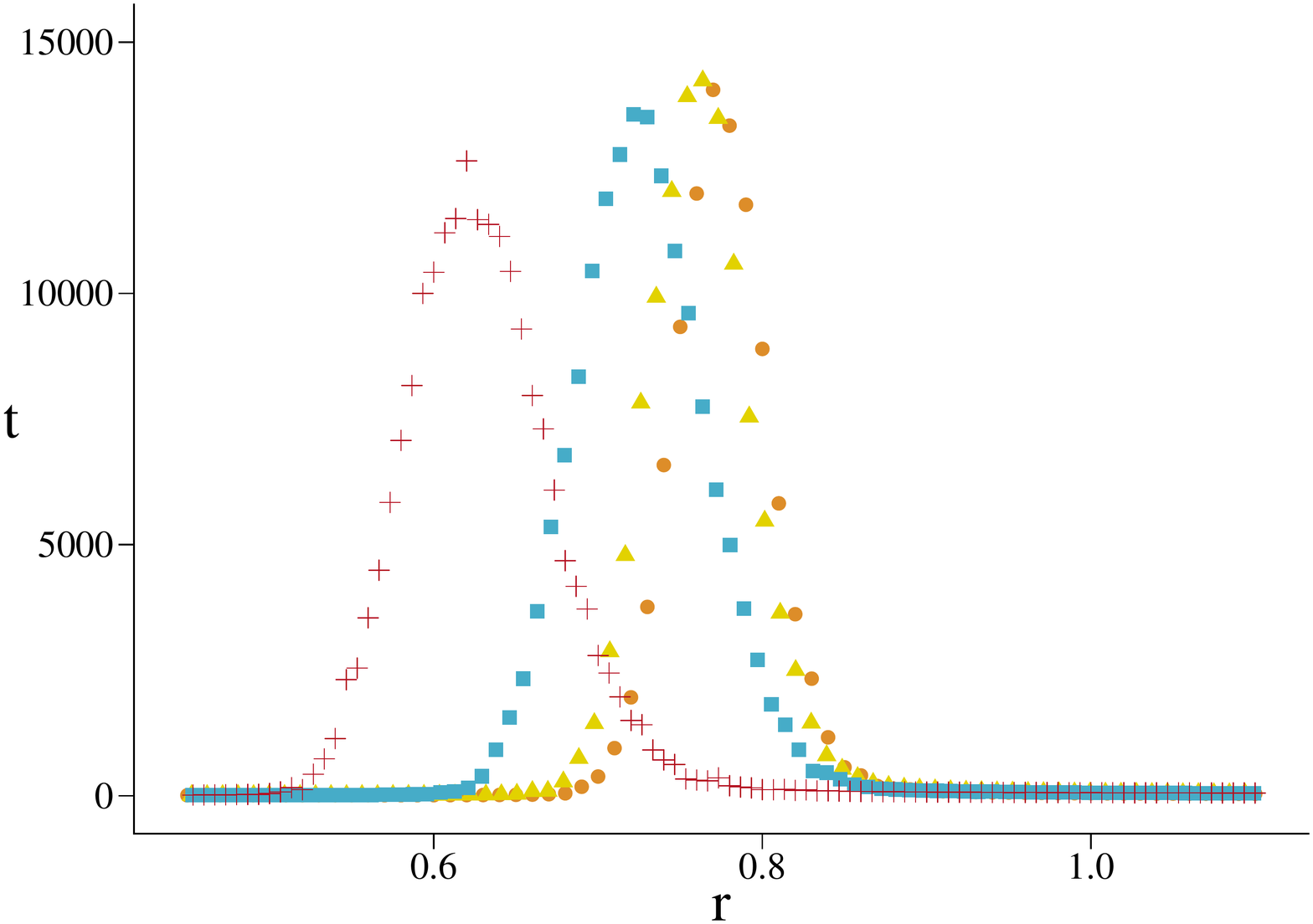}
  \caption{Asymptotic density of cooperators $c$ (top) and convergence time $t$ (bottom) versus $r$, for hypergraphs generated from DM networks with $N=500$ vertices and average degree $\avg{k}=4$. The values of $p$ are specified in the legends; $c_0=0.3$ (left) and $c_0=0.7$ (right), $\beta=1$. The dashed line at $r=1$ indicates the abrupt transition for a well-mixed population.}
  \label{DM_other_c0}
\end{figure}

In Figure~\ref{DM_other_c0} we report the results of the MC simulations performed on hypergraphs derived from DM networks for some values of the initial fraction of cooperators, $c_0$, other than 0.5. The enhanced structural reciprocity when increasing $p$ is confirmed and, additionally, it gets stronger when $c_0$ is increased.

\section{Simulations for the Erd\H os-R\'enyi Model}
\label{sec:appendix_c}

\begin{figure}[tb!]
  \centering
  \includegraphics[width=.46\linewidth]{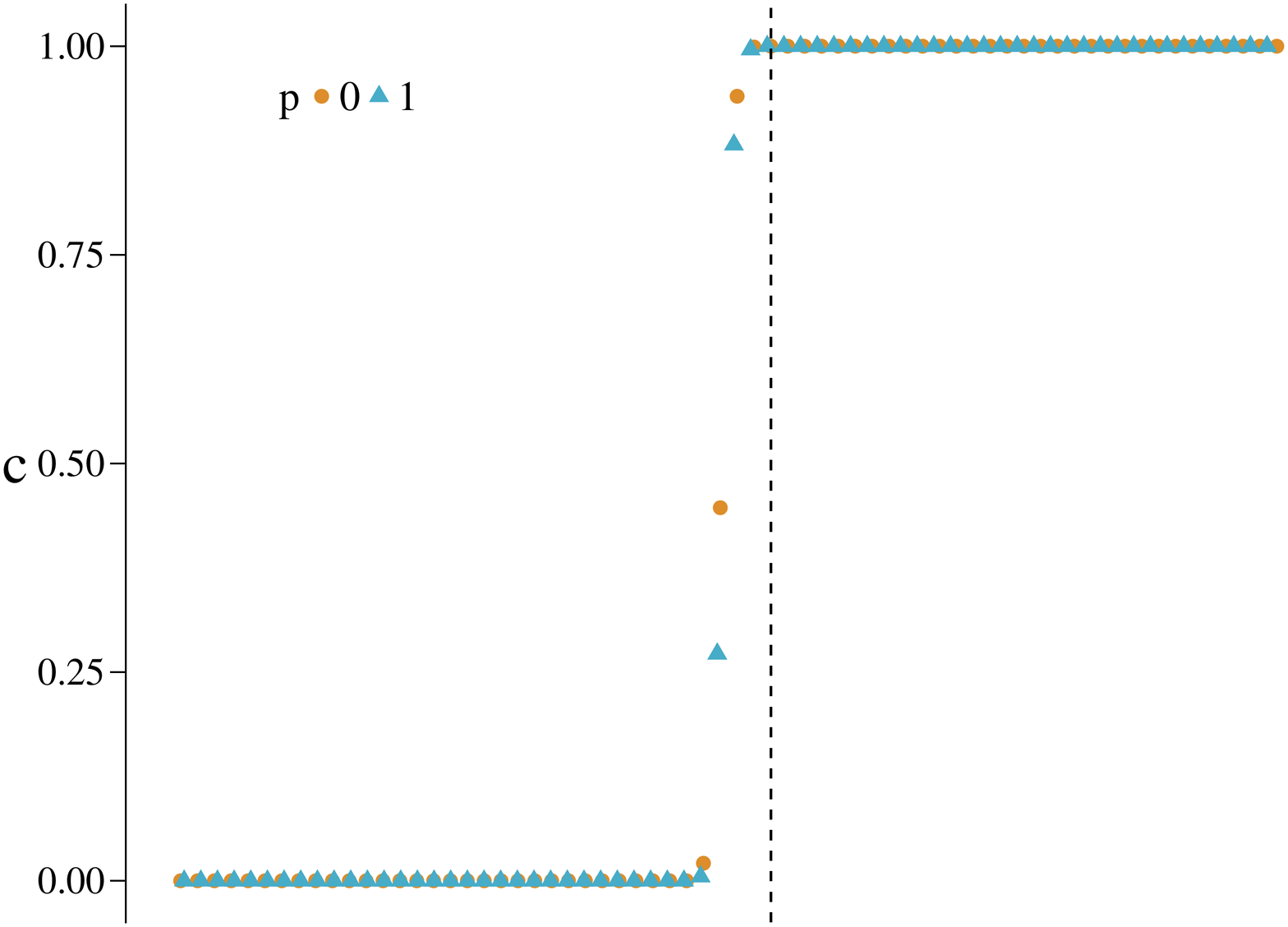}
  \hspace{0.5cm}
  \includegraphics[width=.46\linewidth]{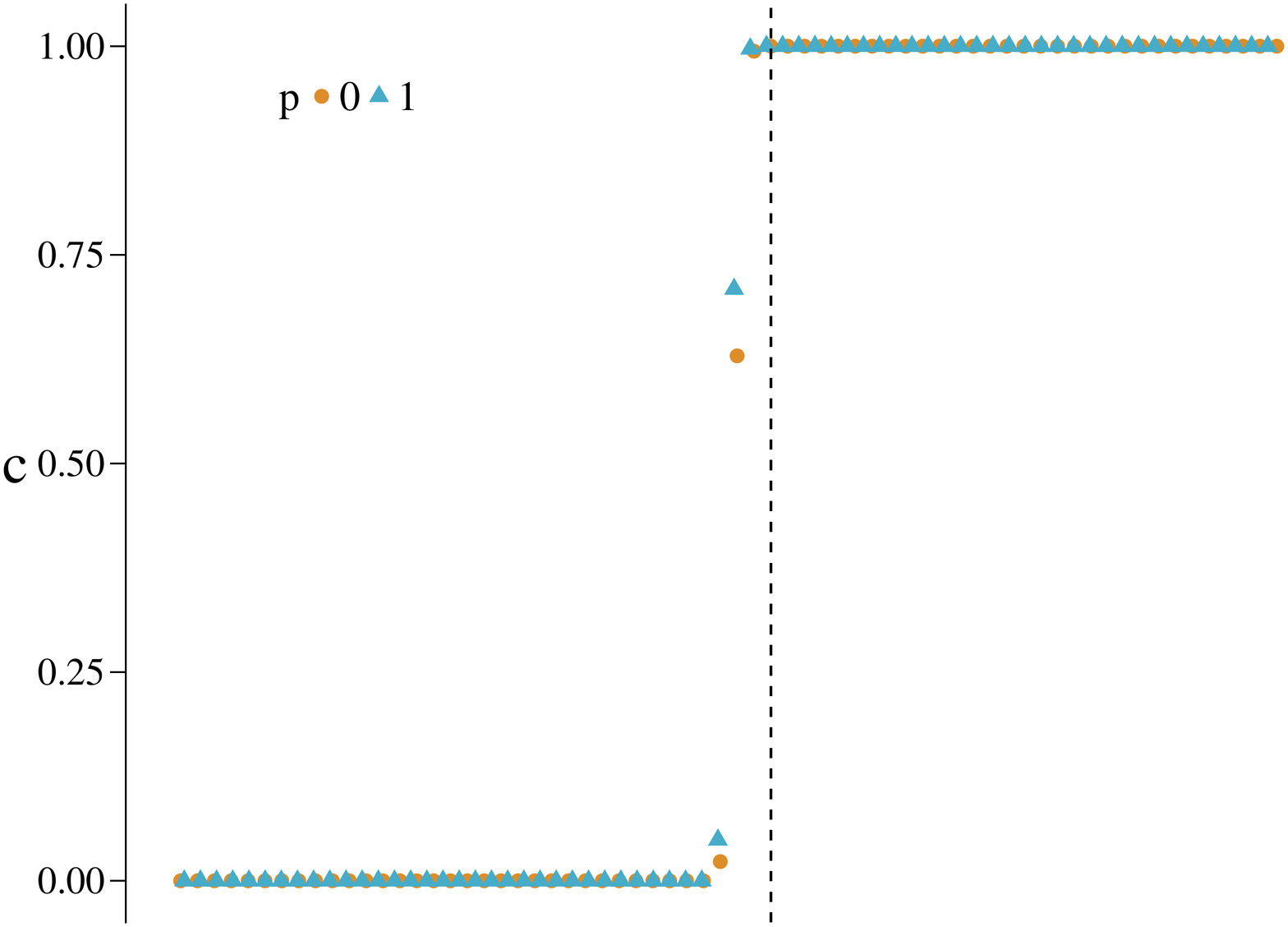}
  \\
  \includegraphics[width=.46\linewidth]{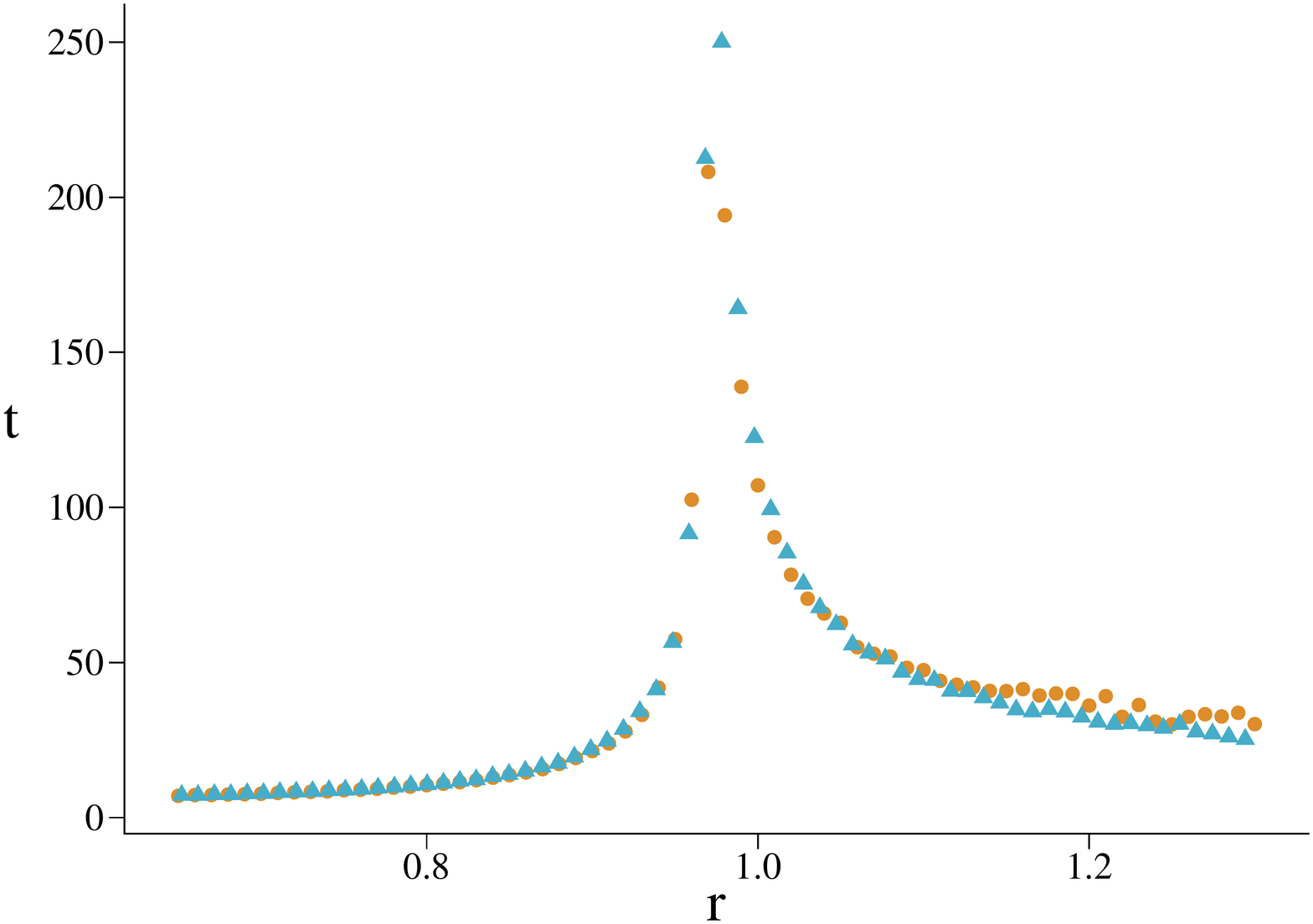}
  \hspace{0.5cm}
  \includegraphics[width=.46\linewidth]{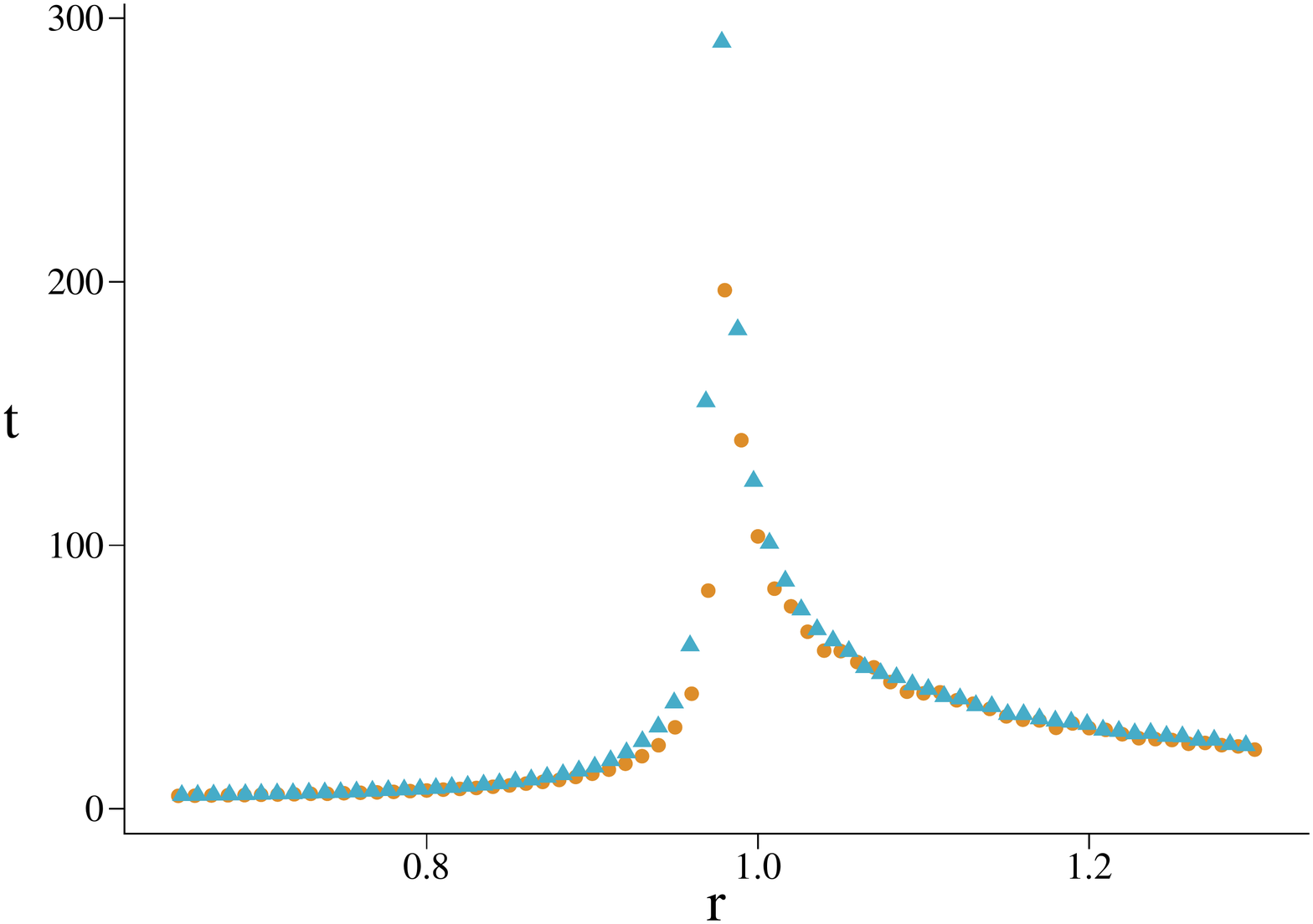}
  \caption{Asymptotic density of cooperation $c$ (top) and convergence time (bottom) versus $r$, for hypergraphs generated from ER networks with $N=500$ vertices, average degree $\langle k\rangle=6$ (left) and $\langle k\rangle=10$ (right). The values of $p$ are specified in the legends; $c_0=0.5$, $\beta=1$. The dashed line at $r=1$ indicates the abrupt transition for a well-mixed population.}
  \label{ER}
\end{figure}

In Figure~\ref{ER} we show the results of the MC simulations performed on hypergraphs stemming from ER networks, taking $\beta=1$ and $c_0=0.5$. The transitions are always sharp and just below $r=1$, as expected for highly homogeneous structures such as ER networks. Accordingly, the denser and less heterogeneous case with $\avg{k}=10$, shows a slightly lower reciprocity. Due to the low fraction of 3-cliques, the transitions are nearly overlapped. Moreover, the average convergence time is always below 300~steps, along with the absence of traps.

\section{On the Role of Cooperator Hubs}
\label{sec:appendix_d}

Equation~(\ref{Hub_threshold}) relies on the comparison between the payoffs of a cooperator hub and any of its defector neighbors at the initial round. For any $r$ greater than the value $r^{\textup{hub}}$ provided by Equation~(\ref{Hub_threshold}), whenever the cooperator hub and any of its defector neighbors match for the strategy update, it is more probable that the defector becomes a cooperator than the other way around. For such values of $r$, the cooperator hub (or more cooperators hub together, as it is likely to be, by construction, in HK and DM networks) can start and sustain the formation of a cooperative community, eventually able to expand over the structure or, at least, to resist to be invaded by any neighboring defective community.

\begin{figure}[tb!]
  \centering
  \includegraphics[width=.97\linewidth]{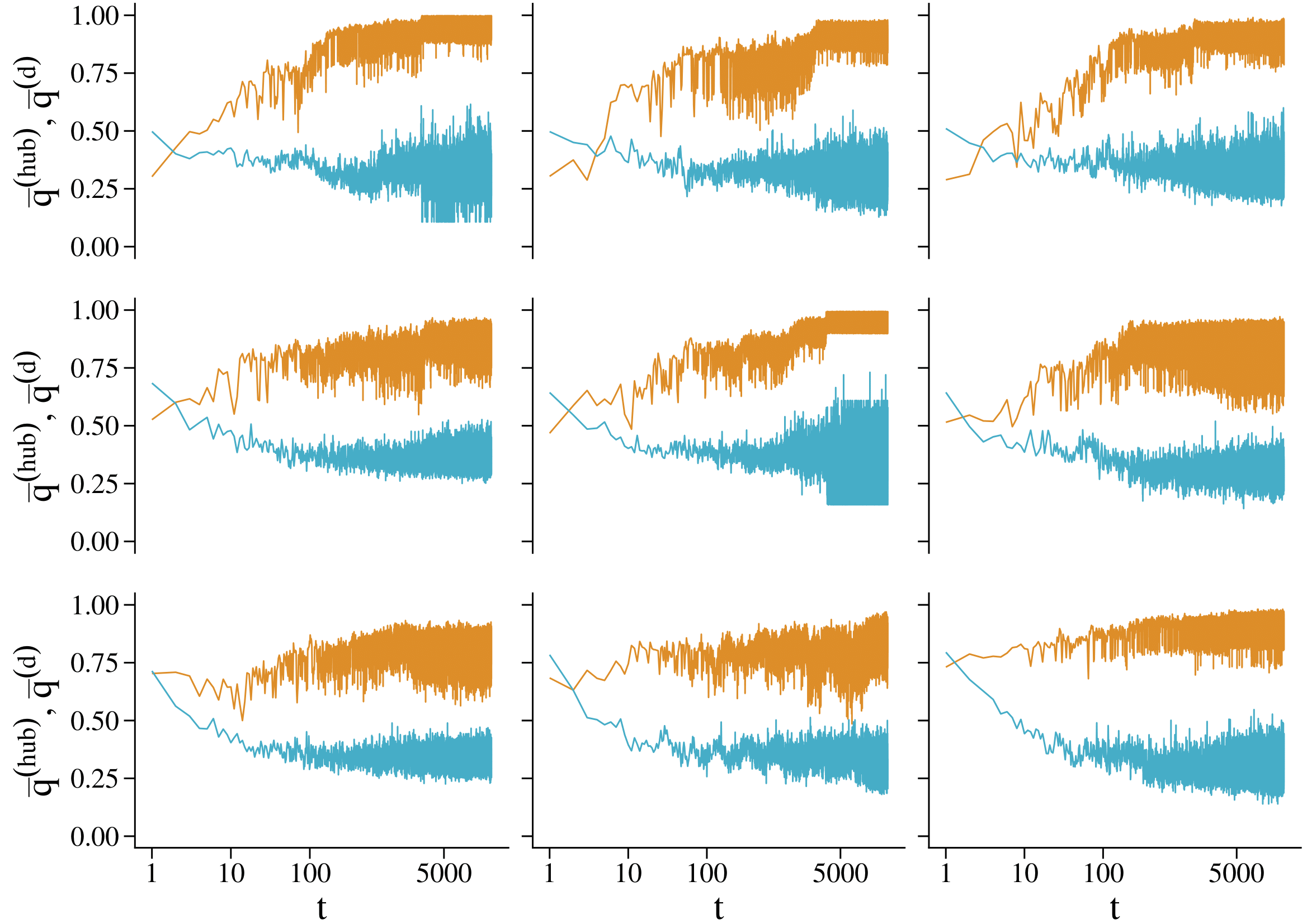}
  \caption{Instances of the temporal evolution of the average fraction of cooperators $\bar{q}^{(hub)}$ in the neighborhood of cooperator hubs (orange), and of the average fraction of cooperators $\bar{q}^{(d)}$ which are neighbors of defector neighbors of cooperator hubs (cyan). First row corresponds to $c_0=0.3$, second row to $c_0=0.5$, third row to $c_0=0.7$. The shown temporal series are obtained for DM networks ($p=0$), taking $\alpha$ around the respective values of $\alpha_{\textup{surv}}$. The hubs are here those vertices with degree above the 98th percentile of the degree distribution. Very similar results are found for HK networks.}
  \label{evo_hubs}
\end{figure}

In Figure~\ref{evo_hubs} we show some temporal evolution of the average fraction of cooperators in the neighborhood of cooperator hubs, $\bar{q}^{(hub)}$, and of cooperator hubs' defector neighbors, $\bar{q}^{(d)}$, for simulations ended with $c_\infty>0$ (not converging), taking $\alpha$ around the respective values of $\alpha_{\textup{surv}}$. It is evident how, independently of $c_0$, $\bar{q}^{(hub)}$ and $\bar{q}^{(d)}$ rapidly reach values around 0.75--0.85 and 0.3--0.35, respectively. Therefore, Equation~(\ref{Hub_threshold}) estimates the conditions for which such temporal evolution (or those ending with $c_\infty = 1$) are likely to start. The prediction gives a clear qualitative justification to the fact that $r$ decreases (the reciprocity improves) with the increasing of both $c_0$ and $s$ (the rank of the hypergraph). The equation gives also a good quantitative prediction for $c_0$ roughly between 0.4 and 0.7. Outside this range, the predicted values of $r$ are, too low for high values of $c_0$, and too high for low values of $c_0$. In other words, the structures provide a lower and a higher reciprocity, respectively, than the one expected by the only comparison of the payoffs at the first rounds. The point is that the prediction made through Equation~(\ref{Hub_threshold}) is mainly based on the average degree disparity among the hubs and their neighbors, while it does not consider the dependence on $c_0$ of the probability that any of those hubs and their neighbors select each other for the strategy update. As long as $c_0$ takes values close enough to $0.5$, the main contribution comes from the degree disparity and Equation~(\ref{Hub_threshold}) works well. However, in asymmetric setups corresponding to low and high values of $c_0$, the approximation used in Equation~(\ref{Hub_threshold}) is no longer sufficient and, consequently, $\alpha_{\textup{th}}^{\textup{hub}}$ does not give an accurate quantitative estimation of $\alpha_{\textup{surv}}$.

\section{On the Local Clustering Coefficient}
\label{sec:appendix_e}

\begin{figure}[tb!]
  \centering
  \includegraphics[width=.46\linewidth]{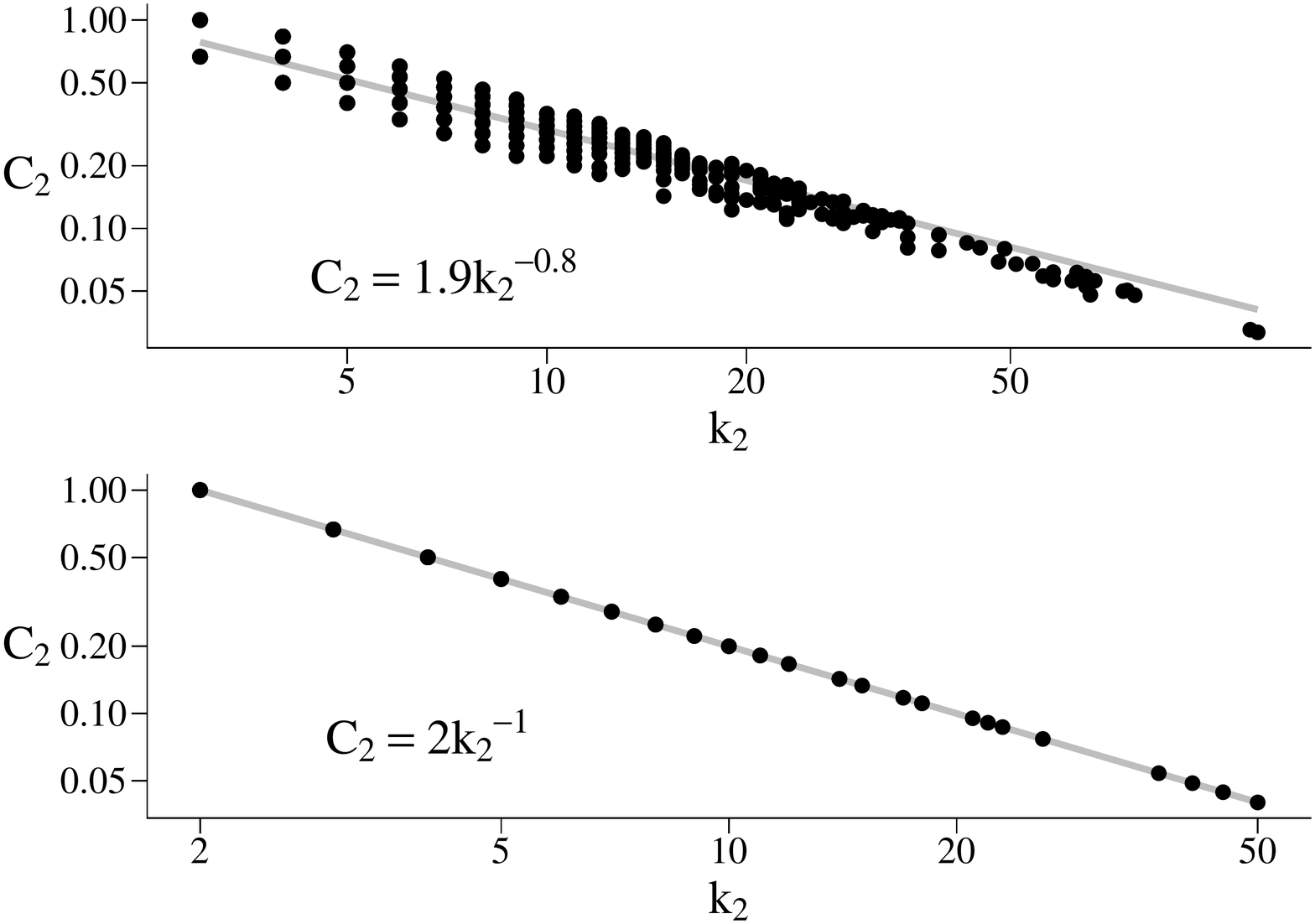}
  \hspace{0.5cm}
  \includegraphics[width=.46\linewidth]{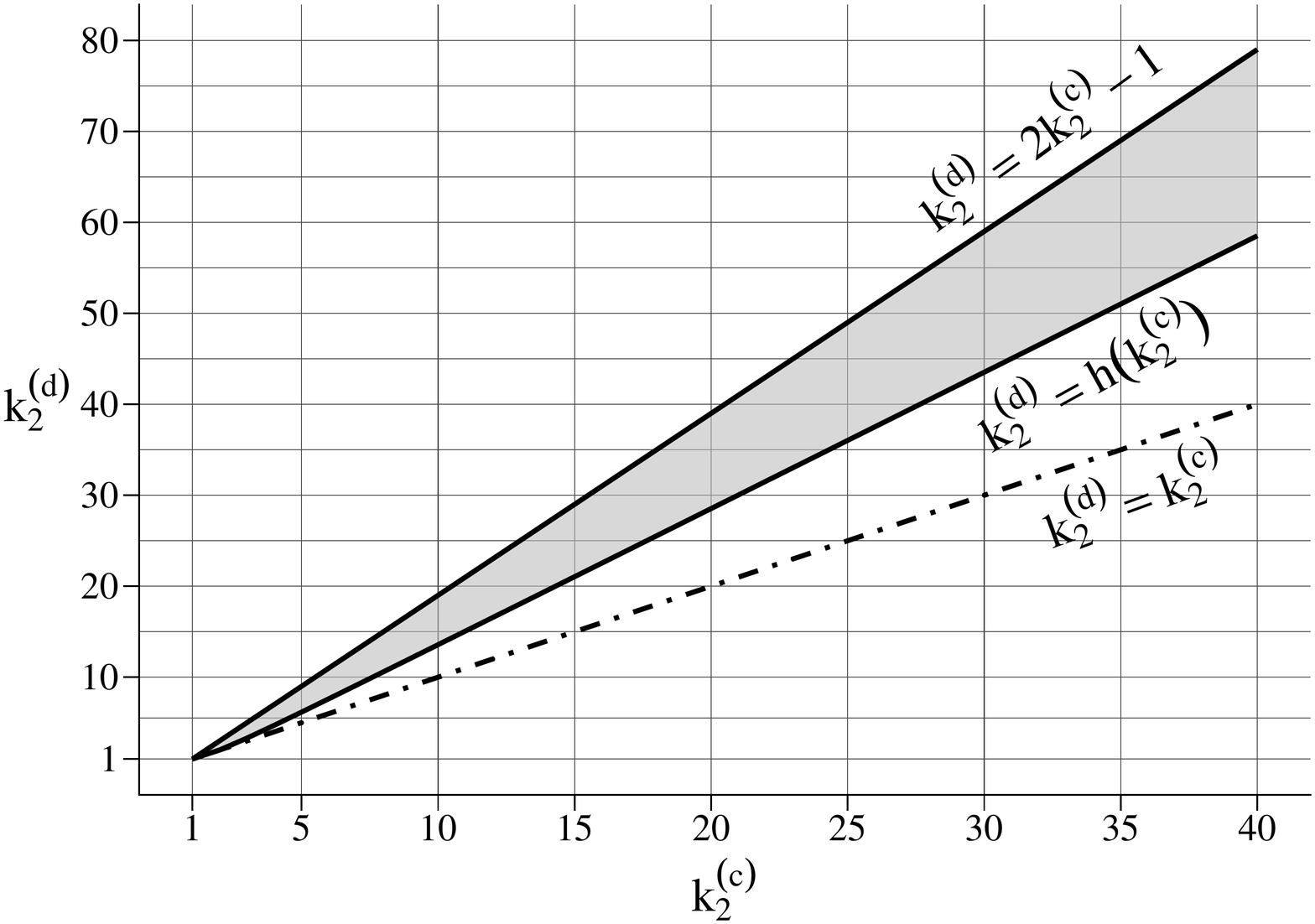}
  \caption{{\em Left}. Local clustering coefficient $C_2$ versus degree $k_2$ for HK (top) and DM networks (bottom). {\em Right}. The shaded area indicates, for a given value of $k_2^{(c)}$, the values of $k_2^{(d)}$ satisfying Equation~(\ref{inequalities_DM}); $h(x)=(3/4)\left[(x-1)+\sqrt{(x-1)^2+(4/3)^2}\right]$.}
  \label{clust_coeff_&_condA6}
\end{figure}

Both HK and DM models grow networks whose local clustering coefficient $C$ scales with the degree $k$ as $C(k)=C(1)k^{-\gamma}$. Using the notation $k_m$ to indicate the $m$-degree, the latter expression reads $C(k_2)=C(1){k_2}^{-\gamma}$. As reported in Figure~\ref{clust_coeff_&_condA6} (left panels), vertices in a DM network follow exactly this law, by construction, with $C(1)=2$ and $\gamma=1$. Vertices in a HK network ($P_t=1$), instead, follow it statistically, with $C(1)\approx 1.9$ and $\gamma\approx 0.8$. Knowing $C(k_2)$, one can express the average number of 3-cliques (or triangles), $\kappa^{(i)}$, that the neighbors of a vertex $i$ share with it, in terms of $k_2$ only. If $\kappa^{(i)}>1$, then some overlap exists among those 3-cliques. We used this quantity in our invasion analysis and it is strictly related with the clustering coefficient $C^{(i)}$ of node $i$. The number $L^{(i)}$ of links among neighbors of vertex $i$ is exactly the number of 3-cliques incident on $i$. This number, if the 3-cliques cover the entire neighborhood of $i$, is the 3-degree $k_{3}^{(i)}$ when we set $p=1$. Then, on one hand,
\begin{equation}
    \kappa^{(i)}=\frac{2{\left.k_3^{(i)}\right|}_{p=1}}{{\left.k_2^{(i)}\right|}_{p=0}}
    \label{kappa}
\end{equation}
on the other,
\begin{equation}
    C^{(i)}=\frac{2{\left.k_3^{(i)}\right|}_{p=1}}{{\left.(k_2^{(i)}(k_2^{(i)}-1))\right|}_{p=0}}
    \label{clustering}
\end{equation}
and the two quantities are related by
\begin{equation}
    \kappa^{(i)}=C^{(i)}{\left.\left(k_2^{(i)}-1\right)\right|}_{p=0}
    \label{kappa_clustering}
\end{equation}
In particular, for DM networks, Equation~(\ref{kappa_clustering}) becomes
\begin{equation}
  \kappa^{(i)} = {\left.2\,\frac{k_2^{(i)}-1}{k_2^{(i)}}\right|}_{p=0}
\end{equation}
and
\begin{equation}
  {\left.k_3^{(i)}\right|}_{p=1} = {\left.(k_2^{(i)}-1)\right|}_{p=0}
\end{equation}
For HK networks ($P_t=1$), instead, we get
\begin{equation}
  \kappa^{(i)} = 1.9\,{\left.\frac{k_2^{(i)}-1}{\left(k_2^{(i)}\right)^{0.8}}\right|}_{p=0}
\end{equation}
and
\begin{equation}
  {\left.k_3^{(i)}\right|}_{p=1} = 0.95\,{\left.\left(k_2^{(i)}-1\right)\left(k_2^{(i)}\right)^{0.2}\right|}_{p=0}
\end{equation}
In the end, this shows that the values of the structural parameters chosen for the computation of Equation~(\ref{def_invasion}) in the main text (Figure~\ref{inv_analysis}) are exact and nearly compatible with DM and HK hypergraphs.

Since the algorithm presented in Section~\ref{sec:methods}, used to generate the hypergraphs, relies on the 2-sections (base networks), Equations~(\ref{kappa})--(\ref{kappa_clustering}) can be generalized to any order. Let us suppose that, from a given 2-section, we can construct uniform hypergraphs with rank up to $m_{\max}$. To generalize the given algorithm, let us define $p_m$, for $m\in\{3,\ldots,m_{\max}\}$, as the fraction of $m$-cliques we transform in hyperedges of degree $m$; $p_3$ corresponds to the $p$ used throughout the text. Whenever we take $p_{\ell}=0$, $\forall \ell\neq m$, we get a $m$-uniform hypergraph for $p_m=1$, and its 2-section for $p_m=0$. Then, given a vertex $i$, the following relations hold:
\begin{align}
  {\left.k_m^{(i)}\right|}_{p_m=1} &= {\left.k_2^{(i)}\right|}_{p_m=0}\,{\left.\frac{\kappa_m^{(i)}}{m-1}\right|}_{p_m=1}
  \\
  {\left.\kappa_m^{(i)}\right|}_{p_m=1} &= {\left.\frac{\left(k_2^{(i)}-1\right)!}{(m-2)!\left(k_2^{(i)}-m+1\right)!}\right|}_{p_m=0}{\left.C_{m-1}^{(i)}\right|}_{p_m=1}
\end{align}
where $C_m^{(i)}$ is the generalized rank-$m$ \textit{local clustering coefficient} of vertex $i$,
\begin{equation}
    C_m^{(i)}=\frac{N_m^{(i)}}{\dbinom{k_2^{(i)}}{m}}
\end{equation}
with $k_2^{(i)}={\left.k_2^{(i)}\right|}_{p_m=0}$. This generalized clustering coefficient is defined as the ratio between the number $N_m^{(i)}$ of $m$-hyperedges among the neighbors of $i$, and the maximum possible number of such $m$-hyperedges. In particular, $C_2$ is the standard local clustering coefficient defined for graphs. Clearly, if $p_{m+1}$ is set to 1, $N_m^{(i)}$ corresponds to $k_{m+1}^{(i)}$. Additionally, $\kappa_m^{(i)}$ is the average number of $m$-cliques (or $m$-hyperedges) that the neighbors of vertex $i$ share with it. Thanks to the above relations, by just knowing how the rank-$m$ local clustering coefficient depends on the 2-degree $k_2$, one is able to compute $\kappa_m$ and $k_m$ for any vertex and, in turn, the correct values of $\xi_m$, see Equation~(\ref{Hub_threshold}). From both HK and DM networks, and a large class of other random models, one can generate uniform hypergraphs with rank not greater than 3.

We now make explicit the form taken by the condition in Equation~(\ref{inequalities}) when $C(k_2)=C(1){k_2}^{-\gamma}$ and, to get an algebraic solution, we consider DM networks where $\gamma =1$. After some algebra, Equation~(\ref{inequalities}) becomes
\begin{equation}
  \frac{3}{4}\left[\left(k_2^{(c)}-1\right)+\sqrt{\left(k_2^{(c)}-1\right)^2+\left(\frac{4}{3}\right)^2}\right]< k_2^{(d)} \leq 2k_2^{(c)}-1
  \label{inequalities_DM}
\end{equation}
Similar expressions follow for HK networks by numerically solving the condition in Equation~(\ref{inequalities}).

In Figure~\ref{clust_coeff_&_condA6} (right panel), the shaded area indicates the region of values satisfying Equation~(\ref{inequalities_DM}). For such values, the competitive dynamics discussed in the main text exists for $p=1$ but not for $p=0$. Taking, for example, $k_2^{(c)}=8$, Equation~(\ref{inequalities_DM}) is fulfilled for $k_2^{(d)}\in\left[11,15\right]$. Note also that, for $k_2^{(d)} > 2k_2^{(c)}-1$ (i.e., the region above the shaded area in Figure~\ref{clust_coeff_&_condA6}), the competitive dynamics also exists for $p=0$. However, due to the linear dependence of Equation~(\ref{payoffs_diff_isolated_def}) with respect to $\alpha$, $\partial(f^{(c)}-f^{(d)})/\partial\alpha<0$ is always more negative for $p=1$ than for $p=0$ whenever $k_2^{(d)} > k_2^{(c)}$, making the competitive dynamics always fiercer in the former case.

\section{Comparing Results from Randomized Networks}
\label{sec:appendix_f}

\begin{figure}[tb!]
\centering
\includegraphics[width=.46\linewidth]{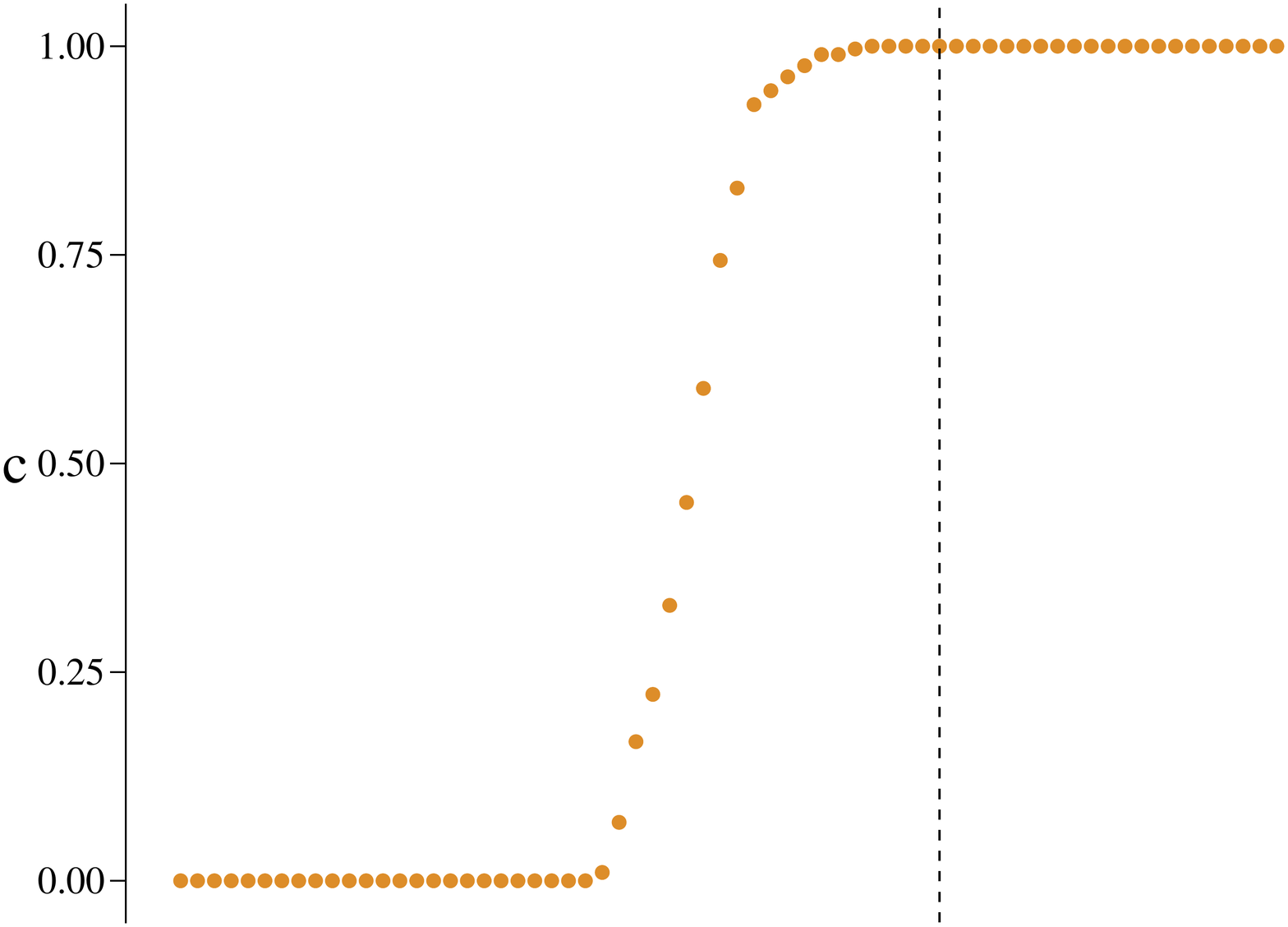}
\hspace{0.5cm}
\includegraphics[width=.46\linewidth]{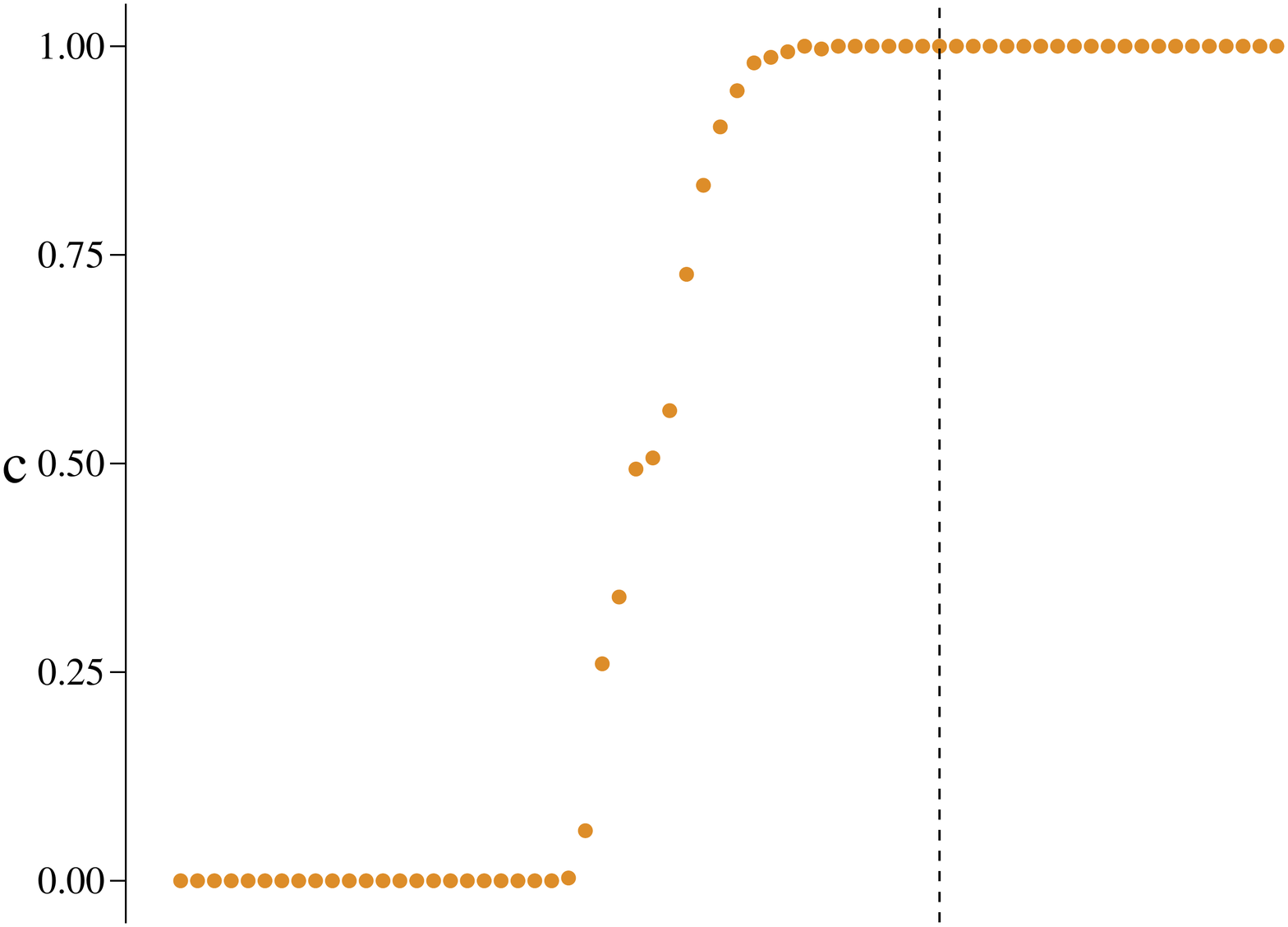}
\\
\includegraphics[width=.46\linewidth]{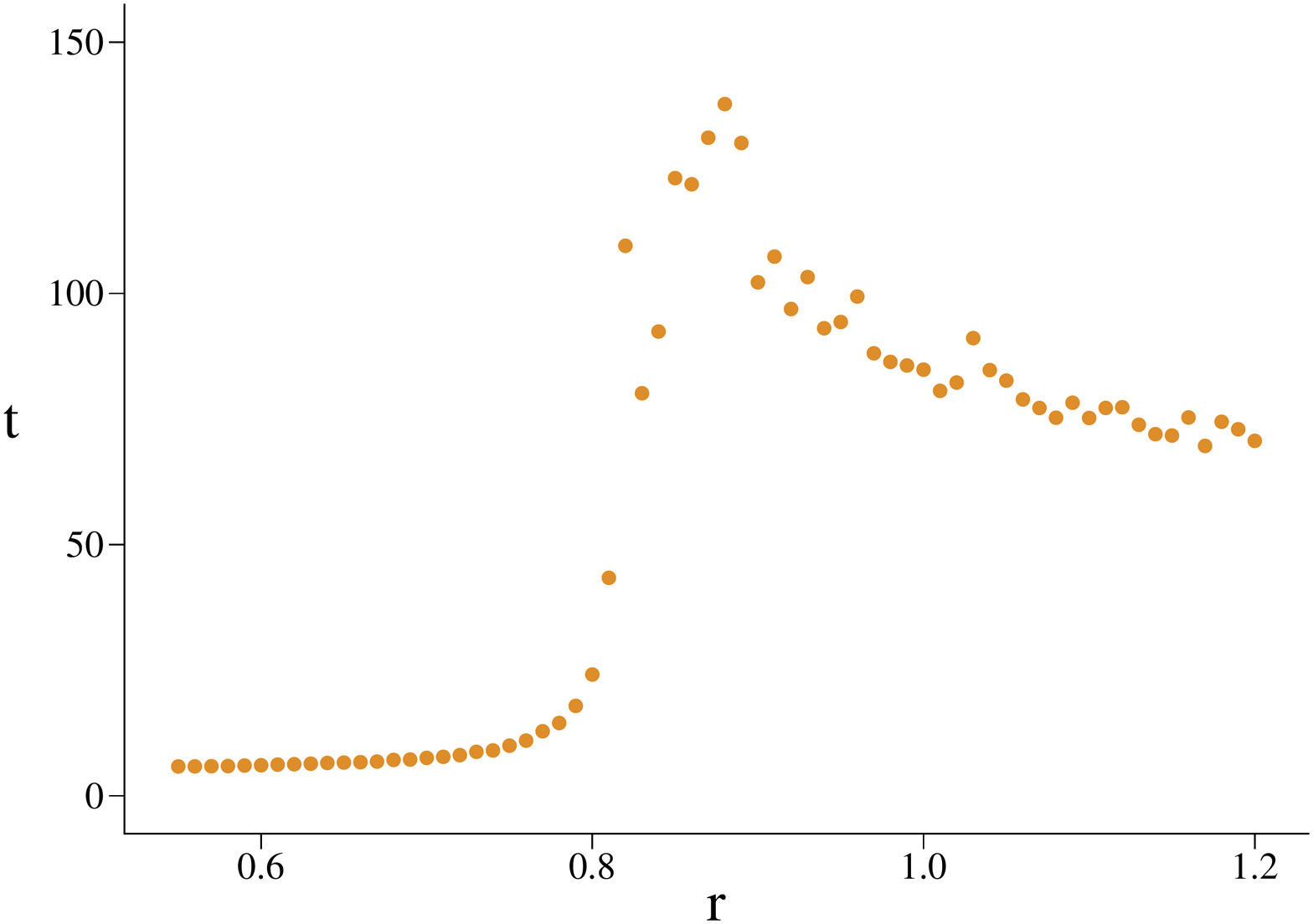}
\hspace{0.5cm}
\includegraphics[width=.46\linewidth]{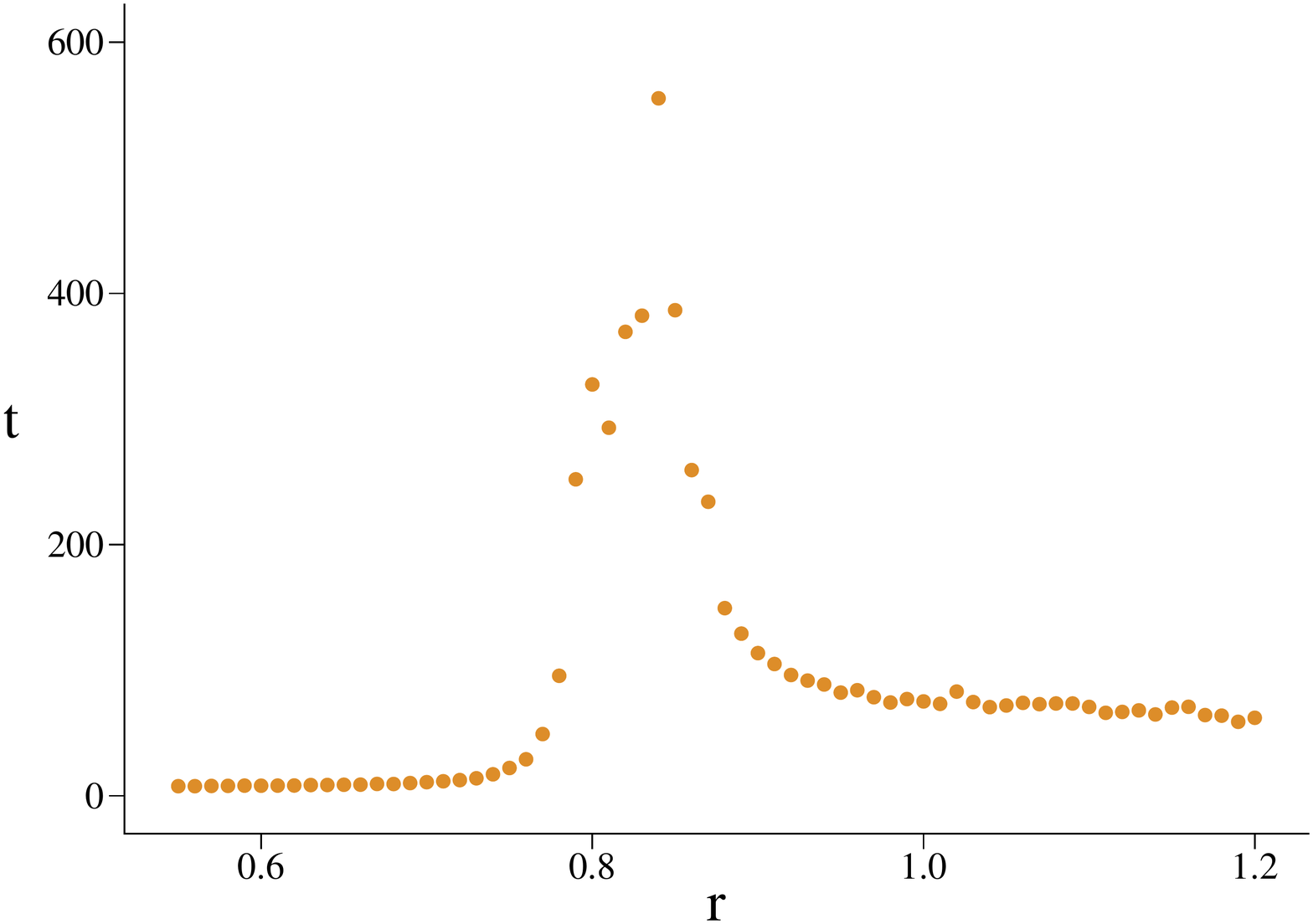}
\caption{Asymptotic density of cooperators $c$ (top) and convergence time $t$ (bottom) versus $r$ for HK (left) and DM (right) networks ($p=0$) with $N=500$ vertices, average degree $\langle k\rangle=4$ and $\langle k\rangle=6$, respectively. The randomization preserves the degree correlations up to the second order, but not the local clustering coefficient; $c_0=0.5$, $\beta=1$. The dashed line at $r=1$ indicates the first-order transition for a well-mixed population.}
\label{randomization_compare}
\end{figure}

We report in Figure~\ref{randomization_compare} the results of the MC simulations performed for networks randomized preserving their second order degree correlations but not fully their local clustering coefficient, indeed lowered to around 0.1. We find $r(0)=0.80$ for HK networks ($P_t=1$), and $r(0)=0.79$ for DM (0.78 and 0.75 without randomization). Additionally, in the critical region, the convergence time is roughly 4~times higher for the latter. This notable discrepancy could be in principle due to both, the higher values of local clustering of DM hypergraphs with respect to HK hypergraphs (0.73 vs.\ 0.41, on average), and to the specific properties of the two models. Since the values of $r(0)$ are very close to the ones found without randomization, this means, on one hand, that the truly important factor responsible for the structural reciprocity is the degree heterogeneity, and on the other, that the better performance shown by DM hypergraphs should be mainly ascribed to their specific structure and scarcely to their higher local clustering.

\reftitle{References}





\end{document}